\pdfoutput=1

\documentclass[useAMS,usenatbib,graphicx]{mn2e}

\usepackage{epsfig}
\usepackage{url}
\usepackage{amssymb}
\usepackage{amsmath}
\usepackage{multicol}
\usepackage{multirow} 
\usepackage{bigdelim}
\usepackage{array}
\usepackage{lscape}

\title[Morphology and environment of galaxies with disc breaks]{Morphology and environment of galaxies with disc breaks in the S$^4$G and NIRS0S}
\author[Laine et al.]
{J. Laine$^{1}$\thanks{E-mail: jarkko.laine@oulu.fi}, E. Laurikainen$^{1,2}$, H. Salo$^{1}$, S. {Comer\'{o}n}$^{1,2}$,  R. J. Buta$^{3}$,
\newauthor 
D. Zaritsky$^{4}$, E. Athanassoula$^{5}$, A. Bosma$^{5}$, J--C. {Mu{\~n}oz--Mateos}$^{6,7}$,
\newauthor 
D. A. Gadotti$^{6}$, J.~ L. Hinz$^{8}$, S. Erroz--Ferrer$^{9,10}$, A. Gil de Paz$^{11}$, 
\newauthor
T. Kim$^{6,7,12}$, K. Men\'{e}ndez--Delmestre$^{13}$, T. Mizusawa$^{7,14}$,
\newauthor 
M.~ W. Regan$^{15}$, M. Seibert$^{16}$, K. Sheth$^{7}$ \\
$^{1}${Astronomy Division, Department of Physics, FIN-90014 University of Oulu, P.O. Box 3000, Oulu, Finland} \\
$^{2}${Finnish Centre of Astronomy with ESO (FINCA), University of Turku, V\"ais\"al\"antie 20, FI-21500, Piikki\"o, Finland} \\
$^{3}${Department of Physics and Astronomy, University of Alabama, Box 870324, Tuscaloosa, AL 35487} \\
$^{4}${University of Arizona, 933 North Cherry Avenue, Tucson, AZ 85721m USA} \\
$^{5}${Aix Marseille Universite, CNRS, LAM (Laboratoire d'Astrophysique de Marseille) UMR 7326, 13388, Marseille, France} \\
$^{6}${European Southern Observatory, Casilla 19001, Santiago 19, Chile} \\
$^{7}${National Radio Astronomy Observatory/NAASC, 520 Edgemont Road, Charlottesville, VA 22903, USA} \\
$^{8}${MMTO, University of Arizona, 933 North Cherry Avenue, Tucson, AZ 85721, USA} \\
$^{9}${Instituto de Astrof\'{i}sica de Canarias, V\'{i}a L\'{a}cteas/n 38205 La Laguna, Spain} \\
$^{10}${Departamento de Astrof\'{i}sica, Universidad de La Laguna, 38206 La Laguna, Spain} \\
$^{11}${Departamento de Astrof\'{i}sica, Universidad Complutense de Madrid, 28040 Madrid, Spain} \\
$^{12}${Astronomy Program, Department of Physics and Astronomy, Seoul National University, Seoul 151-742, Korea} \\
$^{13}${Universidade Federal do Rio de Janeiro, Observat{\'o}rio do Valongo, Ladeira Pedro Ant{\^{o}}nio, 43, CEP 20080-090, Rio de Janeiro, Brazil} \\
$^{14}${Florida Institute of Technology, Melbourne, FL 32901} \\
$^{15}${Space Telescope Science Institute, 3700 San Martin Drive, Baltimore, MD 21218, USA} \\
$^{16}${The Observatories of the Carnegie Institution of Washington, 813 Santa Barbara Street, Pasadena, CA 91101, USA}
}
\begin{document}

\date{Accepted March 2014}

\pagerange{\pageref{firstpage}--\pageref{lastpage}} \pubyear{2014}

\maketitle

\label{firstpage}

\begin{abstract}

We study the surface brightness profiles of disc galaxies in the 3.6 $\mu m$ images from the Spitzer Survey of Stellar Structure in Galaxies (S$^4$G) and $K_{\text{s}}$-band images from the Near Infrared S0-Sa galaxy Survey (NIRS0S). We particularly connect properties of single exponential (type I), downbending double exponential (type II), and upbending double exponential (type III) disc profile types, to structural components of galaxies by using detailed morphological classifications, and size measurements of rings and lenses. We also study how the local environment of the galaxies affects the profile types by calculating parameters describing the environmental density and the tidal interaction strength. We find that in majority of type II profiles the break radius is connected with structural components such as rings, lenses, and spirals. The exponential disc sections of all three profile types, when considered separately, follow the disc scaling relations. However, the outer discs of type II, and the inner discs of type III, are similar in scalelength to the single exponential discs. Although the different profile types have similar mean environmental parameters, the scalelengths of the type III profiles show a positive correlation with the tidal interaction strength.

\end{abstract}


\clearpage
\section{Introduction}

Galaxy discs have been known to follow an exponential decline in the radial surface brightness since the study by \citet{freeman1970}. Disc formation is thought to be closely connected to the initial formation of galaxies, the exponential nature being a result of comparable time-scales for viscous evolution and star formation (e.g. \citealt{lin1987}; \citealt{yoshii1989}). The outermost faint regions of galaxies were first studied in detail in edge-on systems by \citet{vanderkruit1979}, who found that the exponential decay of the surface brightness does not always continue infinitely outwards. Instead a sharp change was found with a much steeper surface brightness decline in the outermost parts of the disc. 

More recently the advance in observations has enabled detailed studies of the faint outer regions of more face-on disc galaxies. Similar, yet not as sharp changes in the surface brightness profiles have been found in many face-on galaxies in the optical (\citealt{erwin2005}; \citealt{pohlen2006}; \citealt{erwin2008}; \citealt{gutierrez2011}), and in the infrared (\citealt{munozmateos2013}). Also, most discs are found to be best described as double exponentials with a change of slope between the two exponential subsections. Discs can thus be divided into three main types (\citealt{pohlen2006}; \citealt{erwin2008}): single exponential discs with no break (type I), discs where the slope is steeper beyond the break (type II), and discs where the outer slope is shallower (type III). The type II breaks in face-on galaxies generally appear much further in than the ``truncations'' in edge-on galaxies (see for example \citealt{martinnavarro2012}), thus most likely representing a different feature.

One of the first explanations for type II breaks in the surface brightness profiles was given by \citet{vanderkruit1987}, relating it to angular momentum conservation during the initial collapse of the gas during the galaxy formation. An alternative explanation involves bars, via the influence of the Outer Lindblad Resonance, which is connected to the formation of outer rings (classified as ``II.o-OLR'', \citealt{pohlen2006}, see also \citealt{munozmateos2013}). Indeed, \citet{erwin2008} noted that in some galaxies with outer rings the break radius of a type II profile is similar to the radius of the outer ring. Furthermore, the presence of a star formation threshold have also been associated with type II profiles in galaxies in which the break radius is larger than the outer ring radius (type ``II.o-CT'', \citealt{pohlen2006}; see also \citealt{schaye2004}; \citealt{elmegreen2006}; \citealt{christlein2010}). Nevertheless, possible connections between breaks and different structural components, such as rings and spirals, have not yet been systematically studied.

Studies of disc breaks in galaxies have become increasingly important with the discovery of stellar migration in discs (e.g. \citealt{sellwood2002}; \citealt{debattista2006}; \citealt{roskar2008a,roskar2008b}; \citealt{schonrich2009a,schonrich2009b}; \citealt{minchev2012}). The idea of migration has changed the paradigm that stars born in the disc do not travel radially far from their place of birth. On the contrary, stars can travel several kiloparsecs radially both inwards and outwards. Evidence of this process has been found in the solar neighbourhood, where the wide metallicity and age distributions of stars \citep{edvardsson1993} can be explained by radial migration \citep{roskar2008b}. The effects of star formation and radial migration can not necessarily be considered separately (e.g. \citealt{roskar2008a}). In these simulations the type II break is caused by a drop of the star formation rate beyond the break due to the reduced amount of cooled gas. However, the outer disc is simultaneously populated by stars radially migrating from the inner disc. In older stellar populations the outer slope is shallower than in younger populations, possibly being a result of more extended radial spreading due to the longer duration of stellar migration (e.g. \citealt{radburnsmith2012}). Colour profiles have also shown that especially for type II profiles the discs become increasingly redder after the break radius (\citealt{azzollini2008}; \citealt{bakos2008}), also consistent with stellar migration. This interpretation is not unique because in the fully cosmological simulations \citet{sanchezblazquez2009} see reddening beyond a break radius in the disc also without stellar migration, and argue that it could simply be due to a change in the star formation rate around the break. In their simulation the presence of stellar migration can smooth the mass profile of the galaxy up to a point where it appears as a single exponential. However, using similar simulations \citet{roskar2010} noted that cosmological simulations can not yet definitely tell the relative roles of radial migration and star formation in the outer regions of galaxies. Alternatively, the larger radial velocity dispersion of old stars (e.g. for solar neighbourhood \citealt{holmberg2009}) could also explain the observed properties of the outer discs up to a point.
        
Type III profiles remain more ambiguous. Sometimes they are associated with an outer spheroidal or halo component in the galaxy, thus not being a disc feature at all (type ``III-s'', \citealt{pohlen2006}; see also \citealt{bakos2012}). \citet{comeron2012} has proposed that $\gtrsim 50 \%$ of type III profiles could also be created by superposition of a thin and thick disc, when the scalelength of the thick disc is larger than that of the thin disc. Extended UV emission has been found in many galaxies beyond the optical disc (\citealt{gildepaz2005}; \citealt{thilker2005}; \citealt{zaritsky2007}). In such cases the increased star formation at the outskirts of the galaxies could give rise to some of the observed type III profiles. Perhaps the most intriguing possibility of type III profile formation comes from the environmental effects. Galaxies live in a hierarchical universe where galaxy mergers are common. These mergers, and also mild gravitational interactions, between galaxies can certainly change the appearance of the involved galaxies. Already in the early simulations of \citet{toomre1972} close encounters between galaxies were shown to significantly perturb the outer discs of the involved galaxies, and as a result tidal tails and bridges formed. They also showed that the masses of the galaxies affect the outcome, and the less massive galaxy is more strongly perturbed. Furthermore, predictions from more recent simulations have shown that type III profiles could be a result of minor mergers (e.g. \citealt{younger2007}; \citealt{laurikainen2001}).

\citet{pohlen2006} made the first attempt to examine the galaxy environments of the different break types by counting the number of neighbouring galaxies from SDSS within 1 Mpc projected radius, for a recession velocity difference to the target galaxy of $|\Delta v|< 350$ km s$^{-1}$, and absolute magnitude of $M_{\text{r'}}< -16$ magnitudes. They concluded that their criteria for the environment were often too harsh to truly characterise it. More recently \citet{maltby2012} compared field and cluster galaxies at higher redshifts ($z_{phot}>0.055$), and found no differences in the break types in different environments. However, they focused only on the outermost disc regions and possibly missed a significant fraction of profile breaks. Therefore, the question of the influence of galaxy environment on disc profile type remains open.

We study the disc and break parameters measured from the radial surface brightness profiles. We aim to systematically associate disc breaks with specific structural components of galaxies, such as rings, lenses, and spirals. In addition, we perform a detailed environmental analysis searching for possible connections among the different disc profile types with the  galaxy density and the presence of nearby perturbers. As a database we use  3.6 $\mu m$ images from the Spitzer Survey of Stellar Structure in Galaxies (S$^4$G, \citealt{sheth2010}) and $K_{\text{s}}$-band images from the Near Infrared S0-Sa galaxy Survey (NIRS0S, \citealt{laurikainen2011}). The 3.6 $\mu m$ and $K_{\text{s}}$-band images are basically free of extinction, particularly at the large disc radii of most interest here. Both bands trace the old stellar population, which is important because of the expected wavelength dependency of the disc scalelengths (\citealt{bakos2008}; \citealt{radburnsmith2012}). In the environmental analysis we use the 2 Micron All Sky Survey (2MASS) Extended Source Catalog (XSC, \citealt{jarrett2000}) and the 2 Micron All Sky Survey Redshift Survey (RSC, \citealt{huchra2012}).
        
The outline of this paper is as follows. In section \ref{sample-selection} we introduce the sample selection criteria for our study, in section \ref{analysis-methods} we describe the data processing, the analysis methods, and the classification of the profile types. In section \ref{env-ana-methods} we describe the environmental study. In sections \ref{results} and \ref{env_effects} we present the main results of the surface brightness profile analysis and the results of the environmental analysis, respectively. The results are discussed in section \ref{discussion}, and summarised in section \ref{sum-conclusion}. The general parameters of the galaxies, including the environmental parameters, are presented in appendix \ref{app:sample} in Table \ref{app:a}, and the parameters of the discs and breaks in Table \ref{app:b}.

\begin{figure*}
  \begin{center}
    \includegraphics[width=0.95\linewidth]{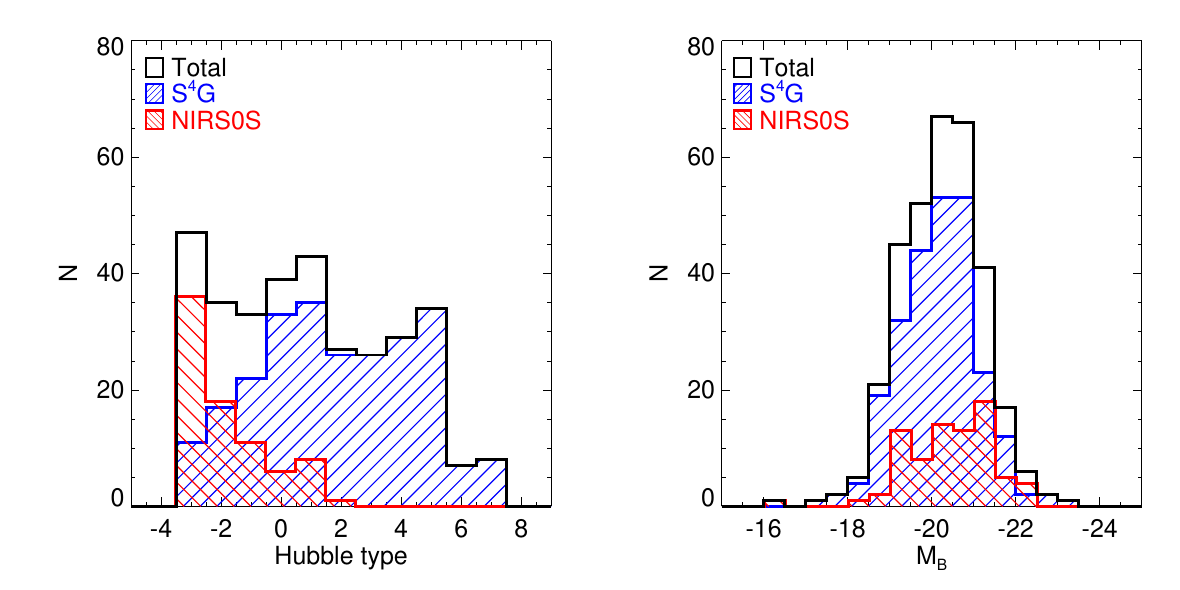}
    \caption{In the \textit{left panel} the Hubble type distribution of the sample galaxies, and in the \textit{right panel} the absolute B-band magnitude distribution of the galaxies is shown.}
    \label{sample_histo}
  \end{center}
\end{figure*}

\section{Sample selection}
\label{sample-selection}

We use the Spitzer Survey of Stellar Structure in Galaxies (S$^4$G, \citealt{sheth2010}) which consists of more than 2300 galaxies observed at 3.6 and 4.5 $\mu m$ wavelengths with the IRAC instrument \citep{fazio2004}. The sample for the survey was selected using values from HyperLeda \citep{paturel2003}, for the radial velocity ($V_{radio} < 3000$ km s$^{-1}$), the total corrected blue magnitude ($m_{Bcorr}<15.5$), the isophotal angular diameter in blue ($D_{25} > 1.0$ arcmin), and the galactic latitude ($|b| > 30^{\circ}$). Many gas poor early-type galaxies do not have radio based measurements of the radial velocities in HyperLeda, and so are not part of the S$^4$G sample. To fill this gap, we include the Near Infrared S0-Sa galaxy Survey (NIRS0S, \citealt{laurikainen2011}) in our study. NIRS0S is a $K_{\text{s}}$-band study and the sample selection is based on the Third Reference Catalogue of Bright Galaxies \citep{devaucouleurs1991} for the morphological types $-3 \le T \le 1$, total magnitude $B_T \le 12.5$ mag, and inclination $i < 65^{\circ}$. Additional galaxies slightly fainter than using the main selection criteria are included in the NIRSOS sample, so that in total the full sample contains 215 galaxies (13 ellipticals, 139 S0s, 30 S0/a, 33 Sa, and one later type galaxy). NIRS0S partly overlaps with the S$^4$G survey, having 93 galaxies in common.
	
To select our sample from these two surveys we use the following selection criteria:
\begin{itemize}
   \item Hubble stage $-3 \le T \le 7$ from Buta et al. (in preparation) for S$^4$G galaxies and from \citet{laurikainen2011} for NIRS0S galaxies,
   \item Galaxy magnitude in $K_{\text{s}}$-band  $m \le 9.5$, taken from 2MASS (isophotal apparent magnitude measured within an elliptical aperture defined at $K_{\text{s}} = 20$  mag arcsec$^{-2}$ isophote \citep{jarrett2000}),
   \item minor/major axis ratio of the disc $b/a > 0.5$.
\end{itemize} 
The very late-type disc galaxies ($T>7$) are excluded because the study of disc structures becomes increasingly harder due to their patchy nature. The above magnitude limit is used because the completeness limit in the 2MASS Redshift Survey \citep{huchra2012} is 11.75 magnitudes in the $K_{\text{s}}$-band. With this selection we are able to search the environments for neighbouring galaxies that are at most two magnitudes fainter than the primary galaxies (see also section \ref{quantifying}). The inclination criterion is applied because we want to restrict to nearly face-on or moderately inclined galaxies, for which the profile shapes can be determined in a reliable manner. For galaxies that appear both in S$^4$G and NIRS0S we use the S$^4$G images because they are deeper. For S$^4$G galaxies we chose to use the 3.6 $\mu m$ images, due to the wavelength range being closer to the $K_{\text{s}}$-band used in NIRS0S.

This results in a sample of 439 galaxies, 336 from S$^4$G and 103 from NIRS0S. We exclude 111 galaxies for which the surface brightness profile analysis was unreliable, due to bright foreground stars in or near the disc, or when there are large gradients in the image. The final sample has 328 galaxies, 248 from S$^4$G and 80 from NIRS0S. The histogram in Figure \ref{sample_histo} displays the final distribution of the Hubble types, and absolute $B$-band magnitudes \citep{paturel2003}. The general properties of the galaxies are listed in appendix Table \ref{app:a}.


\section{Analysis of the surface brightness profiles}
\label{analysis-methods}

\subsection{Initial data processing and analysis}
\label{data-processing}

The data processing and analysis of the S$^4$G images consists of four pipelines\footnote{The images and many data products are publicly available at the IRSA website. \url{http://irsa.ipac.caltech.edu/data/SPITZER/S4G/} } \citep{sheth2010} through which all the galaxies in the survey are processed. The first pipeline (P1) creates the mosaicked science ready images. The second pipeline (P2) uses SExtractor \citep{bertin1996} to create masks for the foreground stars and image artefacts. These masks were also visually inspected and edited if necessary. The third pipeline (P3) performs automated photometry. The basic parameters of the S$^4$G galaxies in our sample (centre, orientation, ellipticity, background level) were obtained from pipeline four performed on the 3.6 $\mu m$ images (P4, Salo et al. in preparation), which produces two dimensional multi-component decompositions of the galaxies. In P4 the background sky level is measured in 10-20 locations outside the galaxy and a mean of the medians is calculated. The standard deviation, $\sigma_{sky}$, of the sky measurements is also calculated. P4 also gives the position angles and inclinations of the discs. We check visually that the orientation parameters used for the disc do not erroneously correspond to large spherical halos. The measurement regions are mostly outside possible outer rings, in which region the discs are likely to be close to axisymmetric. In cases where the galaxy seems to end at an outer ring, the orientation and ellipticity of the outer disc are considered uncertain ($\sim 15 \%$ of the S$^4$G galaxies in the sample, e.g. NGC2859). However, even in these cases the measurements were included in further analysis. Detailed morphological classifications of the S$^4$G galaxies are taken from \citet{buta2010} and Buta et al. (in preparation), which are used to identify the structural components.
                
For the NIRS0S galaxies we use the cleaned and flux calibrated $K_{\text{s}}$-band images \citep{laurikainen2011}. The values of $\sigma_{sky}$, galaxy centres, position angles, disc inclinations, and the morphological classification are from \citet{laurikainen2011}.

\subsection{Radial surface brightness profiles}
\label{fitting}

\begin{figure*}
  \begin{center}
    \includegraphics[width=0.95\linewidth]{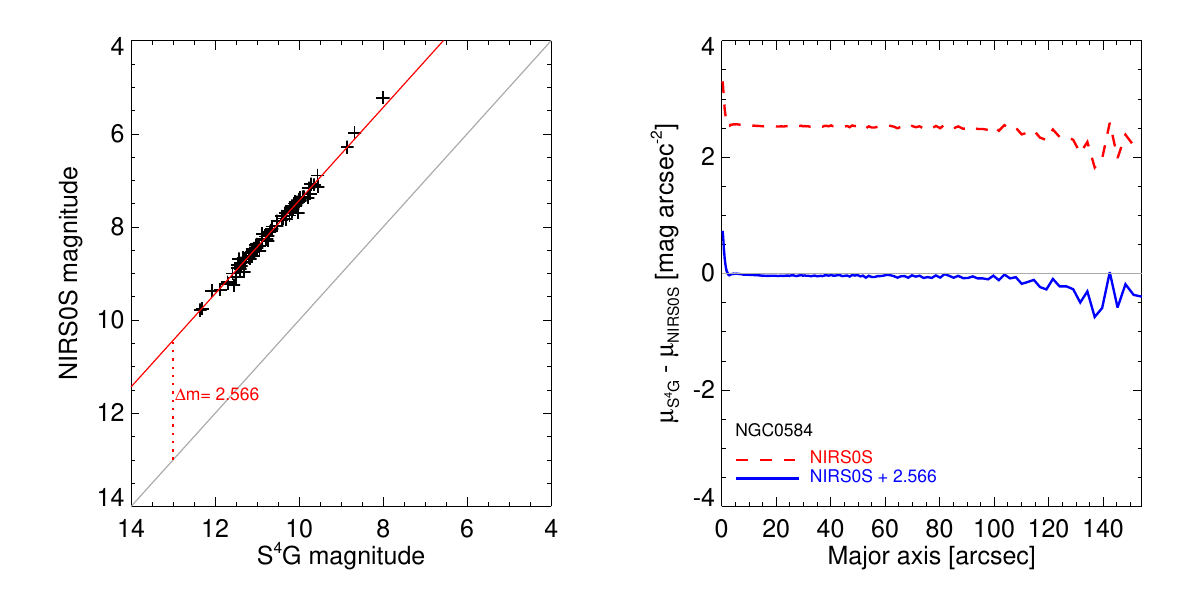}
    \caption{In the \textit{left panel} the aperture magnitudes inside isophote defined by $\mu_{3.6 \, \mu m} = 22.5$ mag arcsec$^{-2}$ are shown for 93 galaxies that are in both S$^4$G and NIRS0S surveys. The median difference in the magnitudes ($\mu^{\text{AB}}_{3.6}$-$K^{\text{Vega}}_{\text{S}}$) is 2.566, which can be used to convert the NIRS0S magnitudes and surface brightnesses to that is used in S$^4$G as shown by the example in the \textit{right panel}.}
    \label{aperture_correction}
  \end{center}
\end{figure*}	

The break features in the galactic discs are studied using azimuthally averaged surface brightness profiles. The profiles are created by running the IRAF\footnote{IRAF is distributed by the National Optical Astronomy Observatories, which are operated by the Association of Universities for Research in Astronomy, Inc., under cooperative agreement with the National Science Foundation.}  task \emph{ellipse}, with the centre, ellipticity, and position angle fixed to the values for the outer discs. For S$^4$G galaxies the surface brightness profiles were converted to AB-system magnitudes in the following manner:
\begin{equation}
  \mu = -2.5 \, \log_{10} \left[ I_{\nu} (\text{MJy  str}^{-1})  \right] + 20.472.
\end{equation}
 The value 20.472 is the zero-point of the flux to magnitude conversion, calculated from the definition of the AB magnitude scale \citep{oke1974}. Aperture correction was not applied to the surface brightness profiles of S$^4$G galaxies (see the IRAC handbook\footnote{ \url{http://irsa.ipac.caltech.edu/data/SPITZER/docs/irac/iracinstrumenthandbook/30/} }, and \citealt{munozmateos2013}) because in our tests the effect was found to be insignificant compared to other error sources. From the flux calibrated NIRS0S images the surface brightness can be calculated as:
\begin{equation}
  \mu = -2.5 \, \log_{10} \left( ADU \right).
\end{equation}
The flux calibration of the NIRS0S images is based on the 2MASS Vega system. In order to convert the NIRS0S surface brightnesses to the AB-system and to evaluate colour differences we took all 93 galaxies common in NIRS0S and S$^4$G. We calculated the total magnitudes of the galaxies inside the elliptical apertures defined by the surface brightness level of $\mu_{3.6 \, \mu m} = 22.5$ mag arcsec$^{-2}$ in the S$^4$G images, for both surveys. The median difference in the aperture magnitudes between NIRS0S and S$^4$G was found to be 2.566. Although the magnitude difference depends also on the 3.6-2.2 $\mu m$ colour of the galaxy, the relation was found to be very linear with colour (see Fig. \ref{aperture_correction} left panel). After adding the conversion factor to the NIRS0S surface brightnesses, the profiles derived from the images of both surveys are very similar (see Fig. \ref{aperture_correction} right panel), except that the S$^4$G images are on average about two magnitudes deeper (see Fig. 4. in \citealt{laurikainen2011}). All surface brightness profiles and magnitudes of the NIRS0S galaxies are converted with this value for the rest of this study. 

The sky measurement uncertainty starts to affect the surface brightness profiles at large radii. The region of the surface brightness profile that falls below the level defined by the standard deviation of the sky ($\sigma_{sky}$) was excluded from the analysis. When it is obvious that the outermost region of the profile is uncertain, the outer limit was taken to be before the limit defined by $\sigma_{sky}$ was reached. The radial surface brightness profiles can be typically followed out to a surface brightness of $\mu = 26.4 \pm 1.2$ mag arcsec$^{-2}$ for the S$^4$G galaxies, and $\mu = 24.7 \pm 0.6$ mag arcsec$^{-2}$ for the NIRS0S galaxies, in both cases expressed in 3.6 $\mu m$ AB magnitudes.

\subsection{Profile functions and the fitting procedure}
\label{proffunc}

To derive the properties of the discs we adopted the methods used in the previous studies of disc breaks (\citealt{erwin2005}; \citealt{pohlen2006}; \citealt{erwin2008}; \citealt{munozmateos2013}). We ignore the inner regions ($ r \, < \, r_{in}$) of the radial surface brightness profiles, where the bulges/bars/inner rings dominate. To recover the properties of the discs, different model functions, depending on the profile shape, are fitted to the radial surface brightness profiles.

The model functions we use to derive the parameters of the discs are the exponential function \citep{freeman1970}, the double-exponential function introduced by \citet{erwin2008}, and the generalization of this introduced by \citet{comeron2012}. The exponential function is
\begin{equation} 
  I(r) = I_0 \exp{\left(- \frac{r}{r_s} \right)},
  \label{function_1}
\end{equation}
where $I_0$ is the surface brightness in the centre of the disc, and $r_s$, the scalelength of the disc. The double-exponential function is the following
\begin{equation}
  I(r) = S \, I_0 \, e^{\frac{-r}{h_i}} \, \left[ 1 + e^{\alpha \left( r - R_{br} \right)} \right]^{\frac{1}{\alpha}  \left( \frac{1}{h_i} - \frac{1}{h_o} \right) }.
  \label{function_2}
\end{equation}
Now $I_0$ is the central surface brightness of the inner exponential section, $h_i$ and $h_o$ are the scalelengths of the inner and outer discs, $R_{br}$ is the break radius, and $\alpha$ is a parameter defining how smooth the break between the inner and outer slopes is. The parameter $S$ is a scaling factor
\begin{equation}
  S^{-1}= \left( 1 + e^{- \alpha R_{br}} \right)^{\frac{1}{\alpha} \left(\frac{1}{h_i} - \frac{1}{h_o}  \right) }.
\end{equation}
Some of the galaxies in our sample have two breaks (i.e. three exponential sections), and for those we use a generalization of function \ref{function_2}
\begin{equation}
  I(r)=S \, I_0 \, e^{- \frac{r}{h_1}} \prod_{i=2}^{i=n} \left\lbrace \left[ 1 + e^{\alpha_{i-1,i}(r-r_{i-1,i})} \right]^{\frac{1}{\alpha_{i-1,i}} \left( \frac{1}{h_{i-1}}-\frac{1}{h_i} \right) } \right\rbrace ,
  \label{function_3}
\end{equation}
where the scaling factor $S$ is
\begin{equation}
  S^{-1} = \prod_{i=2}^{i=n} \left\lbrace \left[ 1 + e^{-\alpha_{i-1,i} \, r_{i-1,i}} \right]^{\frac{1}{\alpha_{i-1,i}} \left( \frac{1}{h_{i-1}} -\frac{1}{h_i} \right)}   \right\rbrace . 
\end{equation} 
Here the parameter $n \, (\ge 2)$ defines the number of exponential sections in the disc, $h_{i-1}$ and $h_{i}$ are the exponential scalelengths inside and outside of the break, $r_{i-1,i}$ is the break radius between the exponential sections with slopes $h_{i-1}$ and $h_i$, and $\alpha_{i-1,i}$ controls the sharpness of that break. The parameter $I_0$ is again the central surface brightness of the innermost section.

These functions were fitted to the surface brightness profiles using the IDL program \emph{mpcurvefit} \citep{markwardt2009}, which uses a non-linear least squares method for fitting. The free parameters in the fits are the disc central surface brightnesses, scalelengths, and the break radii. The parameter describing the smoothness of the break is kept fixed with a value of $\alpha=0.5$, which is a typical value found for this parameter (\citealt{erwin2008}; see also \citealt{comeron2012}).

\begin{figure*}
  \begin{center}
    \includegraphics[width=\linewidth]{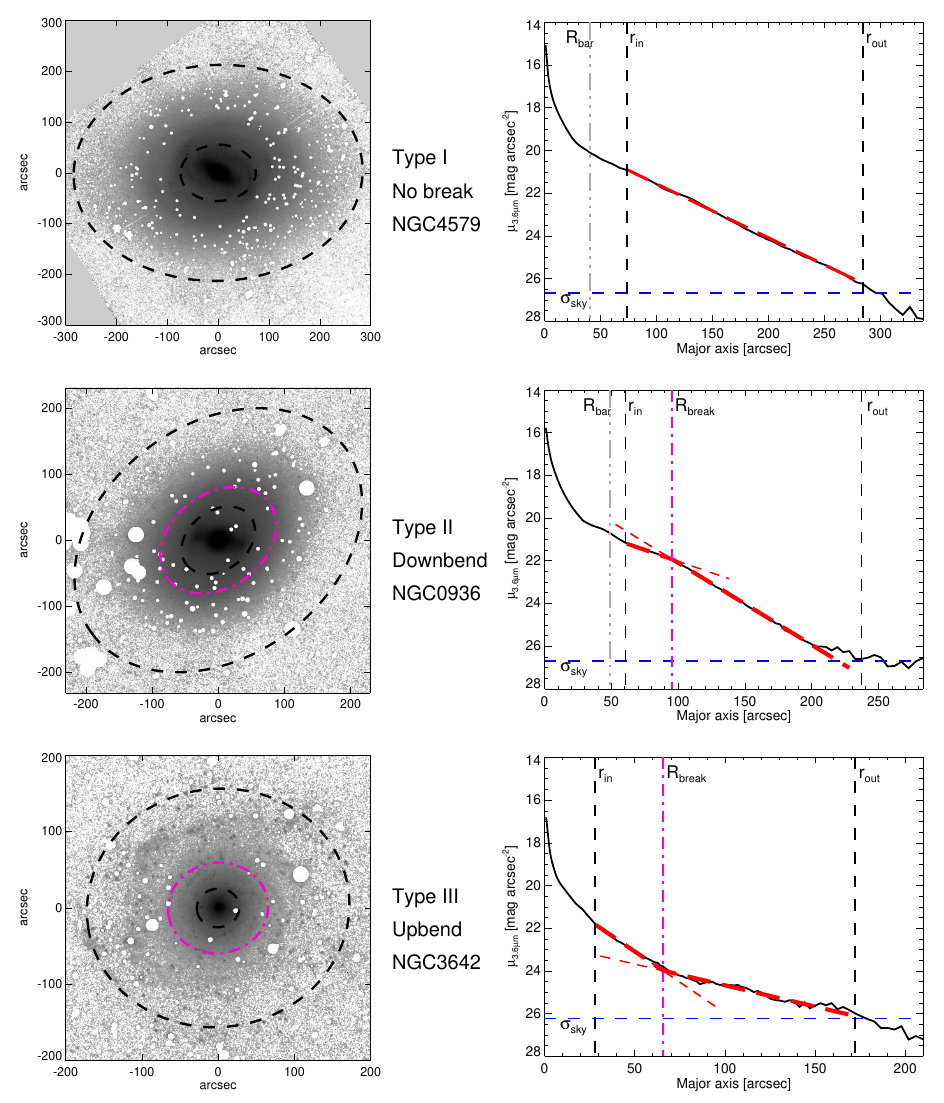}
    \caption{Examples of the different disc types. In the radial surface brightness profiles in the right panels the \textit{horizontal dashed line} presents the sky uncertainty limit, and the \textit{red dashed line} over the surface brightness profile is the fitted profile using the functions explained in the text. The \textit{vertical dashed lines}, and the \textit{dashed ellipses in the images on left}, present the inner- and outer radii of the fitted area. The break radius is also presented in all panels with a \textit{dot-dashed line}. Finally, the visual estimation of the bar radius (if present) is shown in the surface brightness profiles with a \textit{vertical triple-dot dashed line}.}
    \label{prototypes}
  \end{center}
\end{figure*}

The fitting was done in the following steps:
\begin{enumerate}
	\item The inner limit of the fit range, $r_{in}$, is visually identified from the images and the radial surface brightness profiles. The region inside $r_{in}$ is not included in the fit. Figure \ref{prototypes} illustrates how the inner radius is selected for three galaxies. The inner limits are drawn with vertical dashed lines, while the visual estimates of bar lengths (if a bar is present) are indicated with vertical triple-dot dashed lines.
	\item The outer limit of the fit range, $r_{out}$, is defined as the point after which the sky measurement uncertainty starts to dominate. In some cases it was selected visually when it was obvious that there was a contamination in the image, as discussed in section \ref{fitting}. 				
	\item The user identifies the number of sections with exponential slopes. That also determines which model function needs to be used (function \ref{function_1}, \ref{function_2} or \ref{function_3}).
	\item In the case of a single exponential profile the function 3 is fitted to the range defined by [$r_{in}$, $r_{out}$].
	\item In case of a more complicated profile, the initial values of the fit are first estimated by having the user define segments for different exponential slopes in the profile. Then a single exponential profile is fit independently to each of these segments. The values of the central surface brightness and the disc scalelength are then used as initial values, either for the function \ref{function_2} or \ref{function_3}, which is fit to the range defined by [$r_{in}$, $r_{out}$].
\end{enumerate}
		
Possibly the largest source of uncertainty in the disc parameters is due to the selection of the galaxy region where the different functions are fitted (see also \citealt{munozmateos2013}). We follow a Monte Carlo approach to estimate how the fit region selection affects the parameters of the discs. In practice we vary the fit region delimiters ($r_{in}$, $r_{out}$) in an range centred to the original user selected value that has a width of $0.15 \times R_{24}$, and draw 500 new values for the region delimiters using an uniform distribution. The surface brightness profile is then automatically fitted with these region delimiter values, and the standard deviation $\sigma$ of the disc parameters are calculated. The radius defined by the $\mu_{3.6 \, \mu m} = 24$ mag arcsec$^{-2}$ was selected because this level can be reached also with the shallower NIRS0S data. The type of the profile (single exponential, one break, or two breaks) is the same as the user selected, and the initial values for the fit are also the same. The parameters from the fits and the calculated uncertainties ($\pm 1 \sigma$) are shown in the appendix Table \ref{app:b}. The reported uncertainties take into account the possibility of including some of the bulge or the bar region which we try to avoid. In addition, the outer limit of the fit region can extend further out than achieved with the sky uncertainty criteria described above. Therefore possible contribution of the background sky variation is also included in the uncertainties.

\subsection{Classification of the profiles}		

The profiles were classified using the main break types I, II, or III, and in case of multiple breaks, with a combination of types II and III (e.g. \citealt{pohlen2006}; \citealt{erwin2008}), based on the behaviour of the disc scalelengths. Type I is the classical single exponential disc. Type II is a downbending dual-exponential disc, where the outer disc has a lower disc scalelength than the inner disc. Type III is an upbending dual-exponential disc, where the outer disc has a larger scalelength than the inner disc. Examples of the three disc types are shown in figure \ref{prototypes}.

In previous studies, types II  and III have also been divided into several subtypes. For type II discs this is based on the location of the break in relation to the bar length: when the break is at, or inside, the bar radius it is called II.i, and in cases where the break is beyond the bar radius it is called II.o \citep{erwin2005}. We use the II.i class to separate these profiles from pure type I and II.o profiles, which both are intrinsically different from type II.i. We classify the type II.o profiles as type II without separating the subclasses II.o-OLR and II.o-CT. We use similar approach with the type III profiles and do not use the subclasses III-d and III-s. The profile types for each galaxy are presented in appendix Table \ref{app:b}.


\section{Environmental analysis}
\label{env-ana-methods}

For the environmental analysis we use the 2MASS Extended Source Catalog (XSC, \citealt{jarrett2000}) as a basis, due to its completeness in the local Universe. We used objects having apparent $K_{\text{s}}$-band isophotal magnitudes measured in elliptical apertures defined at $K_{\text{s}} = 20$ mag arcsec$^{-2}$ isophote (k\_m\_k20fe), and semi-major axis length corresponding to this isophote (r\_k20fe). All the angular distances between galaxies used in the environmental analysis are calculated from the XSC coordinates. 
	
\subsection{Redshift data}
		
Redshifts for the XSC objects come from the 2MASS Redshift Survey (RSC, \citealt{huchra2012}), and are 97.6\% complete for the XSC galaxies up to $K_{\text{s}} \le 11.75$ mag (44,599 galaxies). The data release also includes redshifts for 196,963 XSC galaxies that are beyond their main catalog limits of the redshift survey (i.e. $K_{\text{s}} > 11.75$ mag, $E(B-V)>1$ mag, or near the Galactic plane). The RSC matches the previous large redshift surveys with the XSC objects, and therefore no additional cross-matching between the XSC and other databases is necessary.

\subsection{Quantifying the environments}
\label{quantifying}

\begin{figure*}
  \begin{center}
    \includegraphics[width=0.95\linewidth]{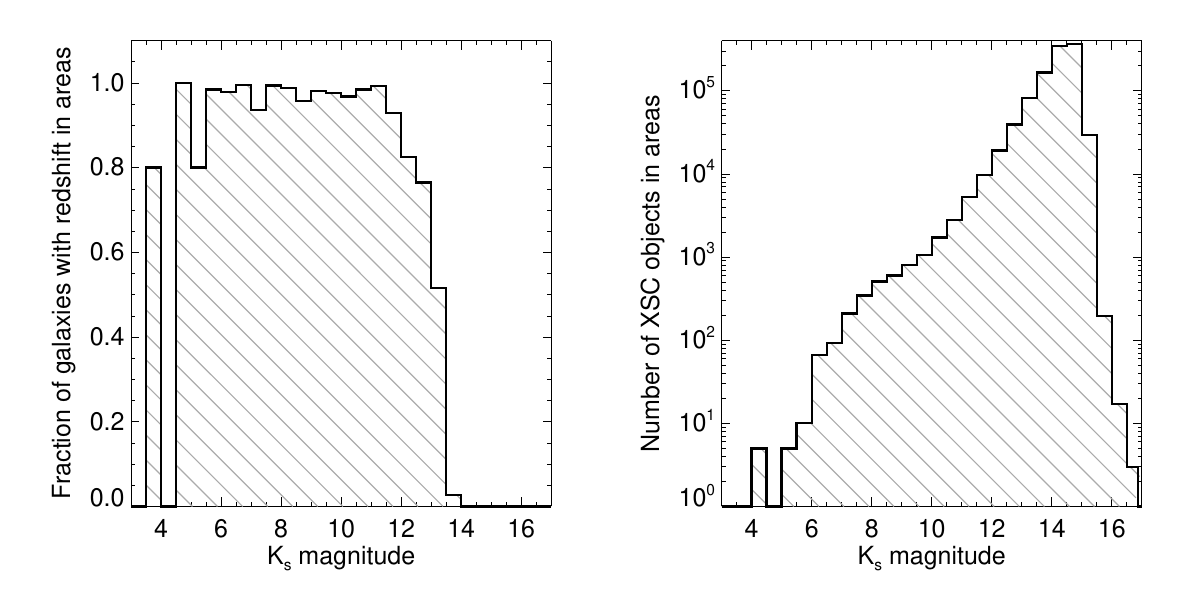}
    \caption{We plot (\textit{left panel}) the redshift completeness in the selected areas around the primary galaxies in 0.5 magnitude bins, and (\textit{right panel}) the number of the galaxies in the same magnitude bins. }
    \label{areas_rs}
  \end{center}
\end{figure*}

We have evaluated galactic environments using two complementary parameters, the projected surface number density of the neighbouring galaxies (e.g. \citealt{dressler1980}; \citealt{cappellari2011}), and the Dahari-parameter \citep{dahari1984}, which estimates the tidal interaction strength between galaxies. These parameters were calculated using galaxies that are found in an area defined by a projected radius of 1 Mpc at the distance of the primary galaxy, and within a recession velocity interval of $\pm 1000$ km s$^{-1}$ around the primary galaxy. In galaxy poor environments the radius of the search area was increased in 1 Mpc steps until there are at least three companion galaxies in the area, while still keeping the velocity interval at $\pm 1000$ km s$^{-1}$. This results in 229 galaxies having 1 Mpc search radius, 45 with 2 Mpc, 16 with 3 Mpc, and 6 with 4 Mpc.		
		
The redshift completeness for the environments were estimated by restricting to XSC galaxies in the search areas. Then, in 0.5 $K_{\text{s}}$-band magnitude bins, the number of galaxies that have a redshift available was divided by the number of all XSC galaxies within the bin. The resulting histograms for the redshift completeness in these bins, as well as the number of XSC galaxies, are shown in Figure \ref{areas_rs}. We confirm that the redshift data is essentially complete for the XSC galaxies up to $K_{\text{s}}$-band magnitude $\sim 12$, as also stated in \citet{huchra2012} for the RSC as a whole. 

The projected surface density of galaxies, using the previously described restriction for the neighbouring galaxies, was calculated with the formula:
\begin{equation} 
  \Sigma_3^A = \log_{10} \left( \frac{N_{gal}}{\pi R_3^2} \right),
  \label{surf_dens}
\end{equation}		
where $N_{gal}=3$ and $R_3$ is the projected distance to the third nearest neighbour galaxy, given in Mpc. We use the logarithm in this parameter because of the large variation of this parameter among the galaxies. The surface density estimated at the radius of the third nearest neighbouring galaxy was selected because it probes the density in small galaxy groups. Estimators using distances to the tenth nearest galaxy work better for surface densities in larger groups or in galaxy clusters, as discussed in \citet{cappellari2011} (e.g. 24 of our sample galaxies are members of Virgo).

We use the Dahari parameter ($Q$, \citealt{dahari1984}) to estimate the gravitational interaction strength. It estimates the tidal force produced by the companion galaxy divided by the internal binding force of the primary galaxy. The Dahari parameter is defined as
\begin{equation} 
Q = \frac{\text{F}_{\text{tidal}}}{\text{F}_{\text{binding}}}= \left( \frac{M_c \, D_p}{S^3} \right) \left( \frac{M_p}{D_p^2} \right)^{-1} = \left( \frac{D_c}{D_p} \right)^{\gamma} \, \left( \frac{D_p}{S} \right)^{3},
  \label{dahari_prop}
\end{equation}
where  $D_p$ and  $D_c$ are the diameters and $M_p$ and  $M_c$ are the masses of the primary and companion galaxies, and $S$ is their projected separation. The value of $\gamma$, the proportionality between the galaxy mass and the radius, is not well known. Here we adopt $\gamma = 1.5$ (\citealt{rubin1982}; \citealt{dahari1984}; \citealt{verley2007}). With this choice the Dahari parameter reduces to
\begin{equation} 
  Q_i \equiv \frac{\left( D_p \, D_c \right)^{1.5}}{S^3}.
  \label{dahari_eq}
\end{equation}
The values for the galaxy diameters were collected from the XSC using two times the semi-major axis length of the ellipse at the isophote of $K_{\text{s}}=20$ mag arcsec$^{-2}$ (r\_k20fe). Following \citet{verley2007} we take $Q$ as the logarithm of the sum of all $Q_i$
\begin{equation} 
  Q = \log_{10} \left( \sum_{i=1}^n  Q_i \right),
  \label{dahari_log}
\end{equation}
where $n$ is the number of galaxies that are found in the search area with recession velocity of $\pm 1000$ km s$^{-1}$ around the primary galaxy. The environmental parameters for individual galaxies are listed in appendix Table \ref{app:a}.


\section{Properties of the breaks and discs}
\label{results}

\subsection{Overall statistics of the breaks}
\label{overall_stat}

\begin{table}
  \centering
    \caption{Statistics of the break types in the surface brightness profiles, compared with those from \citet{gutierrez2011}. We also present the fractions of the profile types separately in barred and non-barred galaxies. We also give the bar fractions for each profile type. The uncertainties are calculated using binomial statistics, and denote the $\pm 1 \sigma$ uncertainties. The overall percentages of the profile types are above 100\% due to seven galaxies that have two breaks in their discs.}
  \begin{tabular}{@{}lrr|r}
    \multicolumn{3}{l}{All galaxies (328)} & \citet{gutierrez2011} (183) \\
    \hline 
    Type & Fraction & N &  Fraction \\
    \hline
    Type I & $32 \% \pm 3 \%$ & 104 & $21 \% \pm 3 \%$ \\
    Type II.i & $7 \% \pm 2 \%$ & 24 & \hspace{-5pt} \rdelim\{{2}{15.3mm}[{$50 \% \pm 4 \%$}]\\
    Type II & $42  \% \pm 3 \%$ & 138 &  \\
    Type III & $21 \% \pm 2 \%$ & 69 & $38 \% \pm 4 \%$ \\
    \hline \\
    \multicolumn{4}{l}{Barred galaxies (204)} \\
    \hline 
    Type & Fraction & N & Bar fraction \\
    \hline
    Type I & $25 \% \pm 3 \%$ & 51 & $49 \% \pm 5 \%$  \\
    Type II.i & $12 \% \pm 2 \%$ & 24 & $100 \% \pm 0 \%$\\
    Type II & $48 \% \pm 4 \%$ & 100 &  $72 \% \pm 4 \%$ \\
    Type III & $16 \% \pm 3 \%$ & 33 & $48 \% \pm 6 \%$ \\
    \hline \\
    \multicolumn{4}{l}{Non-barred galaxies (124)} \\	
    \hline 
    Type & Fraction & N & \\
    \hline
    Type I & $42 \% \pm 5 \%$ & 53 &  \\
    Type II.i & $0 \% \pm 0 \%$ & 0 & \\
    Type II & $30 \% \pm 4 \%$ & 38 &   \\
    Type III & $28 \% \pm 4 \%$ & 36 &  \\
  \hline 
  \end{tabular}
  \label{overall-stat}
\end{table}

In our sample, discs with a single exponential slope appear approximately one third of the time ($32 \% \pm 3 \%$, see Table \ref{overall-stat}). The most common disc type is the type II ($42 \% \pm 3\%$). The type II.i profiles, in which the break is seen inside or at the bar radius, are rare ($7 \% \pm 2 \%$). The type III profiles are the least common of the main profile types ($21 \% \pm 2\%$). Seven of the galaxies have two breaks (II+III or III+II). These galaxies were counted twice, and explain why the total percentage exceeds 100\% in Table \ref{overall-stat}.
               
\begin{figure*}
  \begin{center}
    \includegraphics[width=0.95\linewidth]{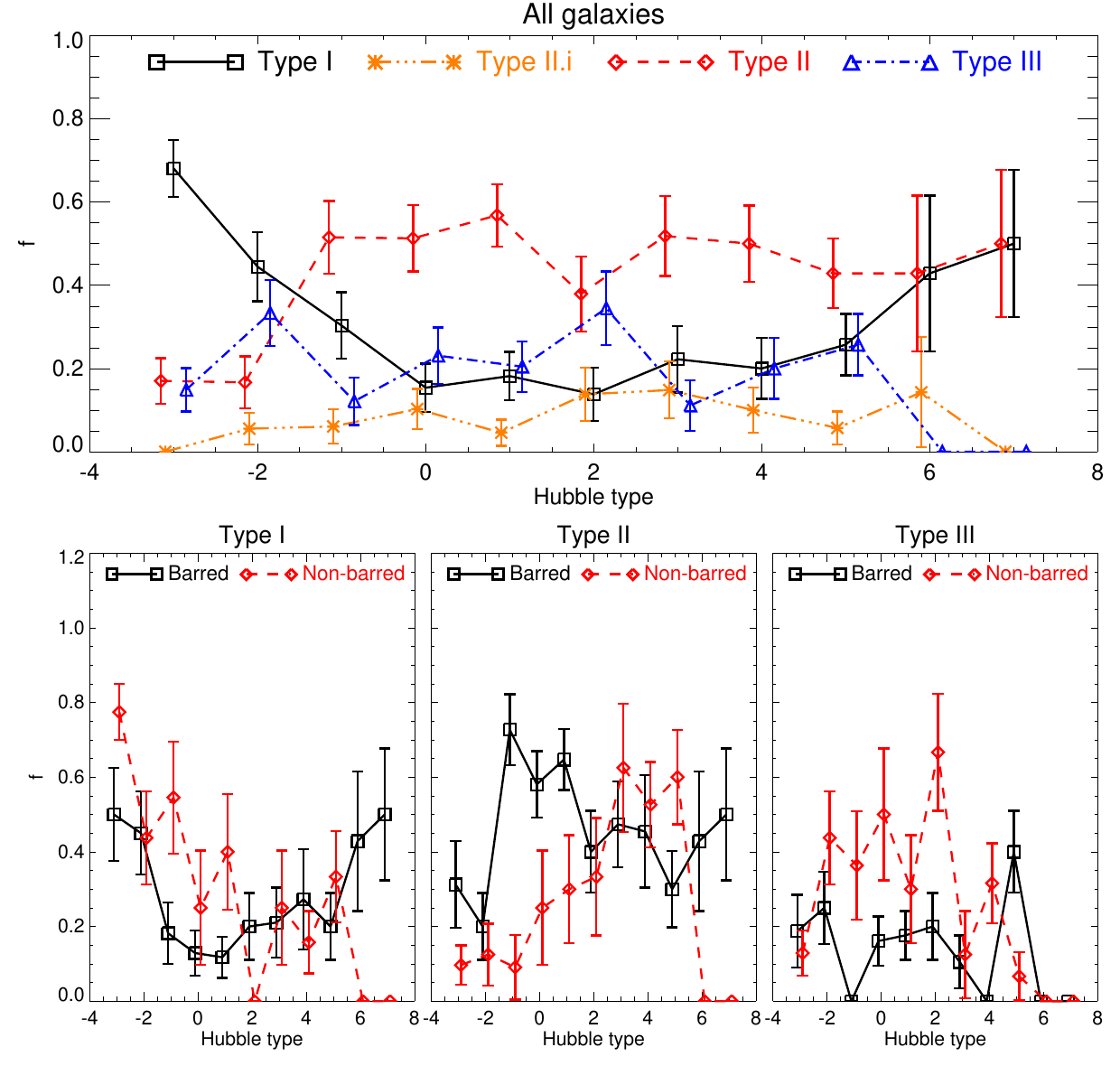}
    \caption{The distribution of the disc-profile types among morphological types. Error bars denote $\pm 1 \sigma$ errors, calculated with the binomial formula $\Delta f = \sqrt{f(1-f)/N}$, where $N$ is the total number of galaxies in the bin. Small offsets are applied to the Hubble type values in the plots for clarity.}
    \label{break_distribution_simple}
  \end{center}
\end{figure*}

In Table \ref{overall-stat} we compare the fractions of the main profile types with those obtained by \citet{gutierrez2011} at optical wavelengths. Clearly, in our sample the fraction of type I profiles is higher, and the fraction of type III profiles is lower. Type II.i and III profiles show no connection with Hubble type (Fig. \ref{break_distribution_simple}, upper panel). Type I profiles are more common in early-type disc galaxies ($T < -1$), and again show a slight increase among the late-type disc galaxies ($T > 5$). Type II profiles are found in about half of the galaxies in each bin, except in the earliest types ($T < -1$). 

The fraction of barred galaxies in our sample $\sim 62 \%$, based on the morphological classification by Buta et al. (in preparation) for S$^4$G and by \citet{laurikainen2011} for NIRS0S, is the same as the fraction typically found in the local universe ($\sim 60-70 \%$ , \citealt{eskridge2000}; \citealt{whyte2002}; \citealt{laurikainen2004}; \citealt{menendezdelmestre2007}; \citealt{marinova2007}). When the bar fractions of the different profile types are examined (Table \ref{overall-stat}) we see that type I and III profiles are equally often found in barred and non-barred galaxies. However, the type II profiles are more common among the barred galaxies (bar fraction is $72 \% \pm 4 \%$). Due to our classification criteria all type II.i profiles have bars. 

In the lower panels of Figure \ref{break_distribution_simple} we show the fractions of the main profile types (I, II, and III) in barred and non-barred galaxies, as a function of Hubble type. It appears that all late-type galaxies in our sample with $T \ge 6$ are barred. Types I and III (Fig. \ref{break_distribution_simple}, lower left and right panels respectively) in barred and non-barred galaxies show similar behaviour. Type II profiles  (Fig.\ref{break_distribution_simple}, lower middle panel) are more common in barred galaxies, especially in Hubble types $T < 3$. In the non-barred galaxies they are rarer in early-type galaxies, the fraction gradually rising for Hubble types $T > 0$.
                              
\subsection{Connection with the structural components}
\label{connections}

\begin{figure*}
  \begin{center}
    \includegraphics[width=0.92\linewidth]{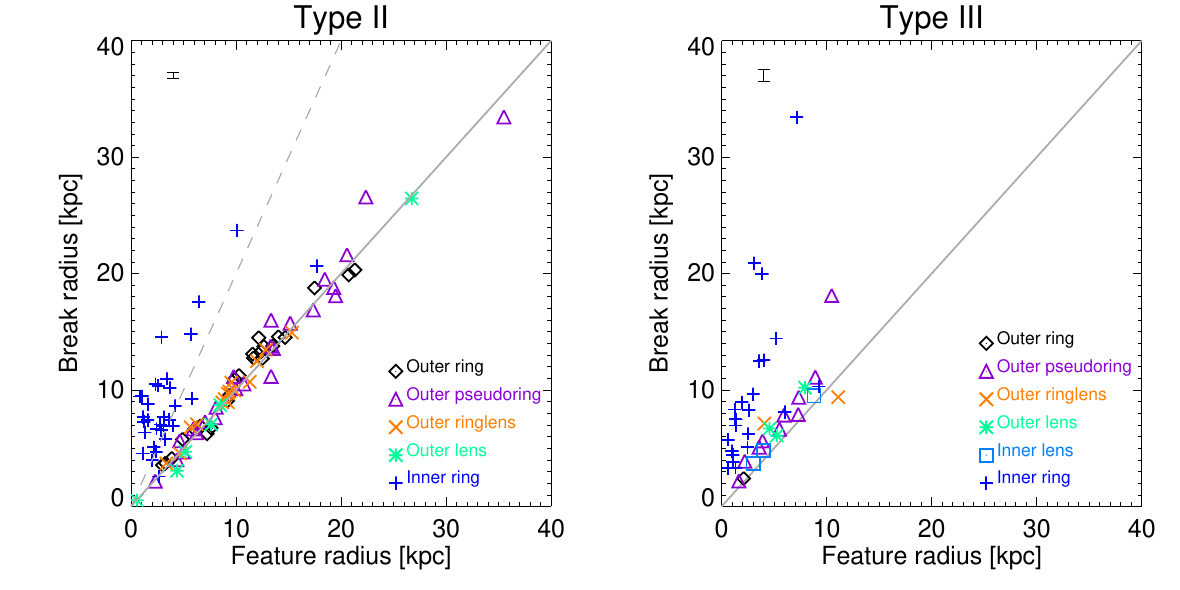}
    \caption{The radii of the closest feature to the breaks is shown against the break radii. Feature dimensions are from \citet{comeron2014} and \citet{laurikainen2011}. Type II breaks are shown at the \textit{left}, and type III breaks at the \textit{right}. The \textit{solid lines} show when the break radius is the same as the feature radius. The \textit{dashed line} in the left panel shows a simple linear fit to the inner rings, with a slope of 2. The error bars in the upper left corners of the panels show the mean uncertainties of the break radius, and they are not large enough to significantly affect the similarity between the break radius and the feature dimension.}
    \label{rings_breaks}
   
    \includegraphics[width=0.92\linewidth]{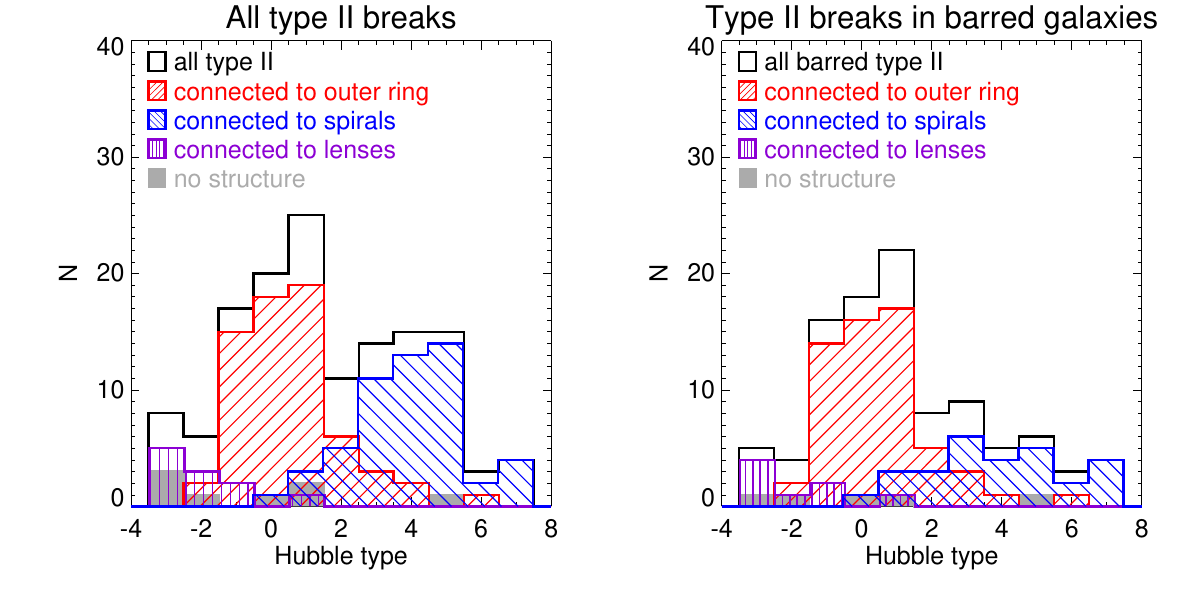}
    \caption{The distribution of type II breaks among morphological types, and the connection with lenses, outer rings, and spiral structures. The galaxies in which no structure could be associated with the break are also shown. In the \textit{left panel} all type II breaks are shown, and in the \textit{right panel} only type II profiles found in barred galaxies are shown.}
    \label{rings_breaks_histo}
  \end{center}
\end{figure*}

\begin{figure*}
  \begin{center}
    \includegraphics[width=\linewidth]{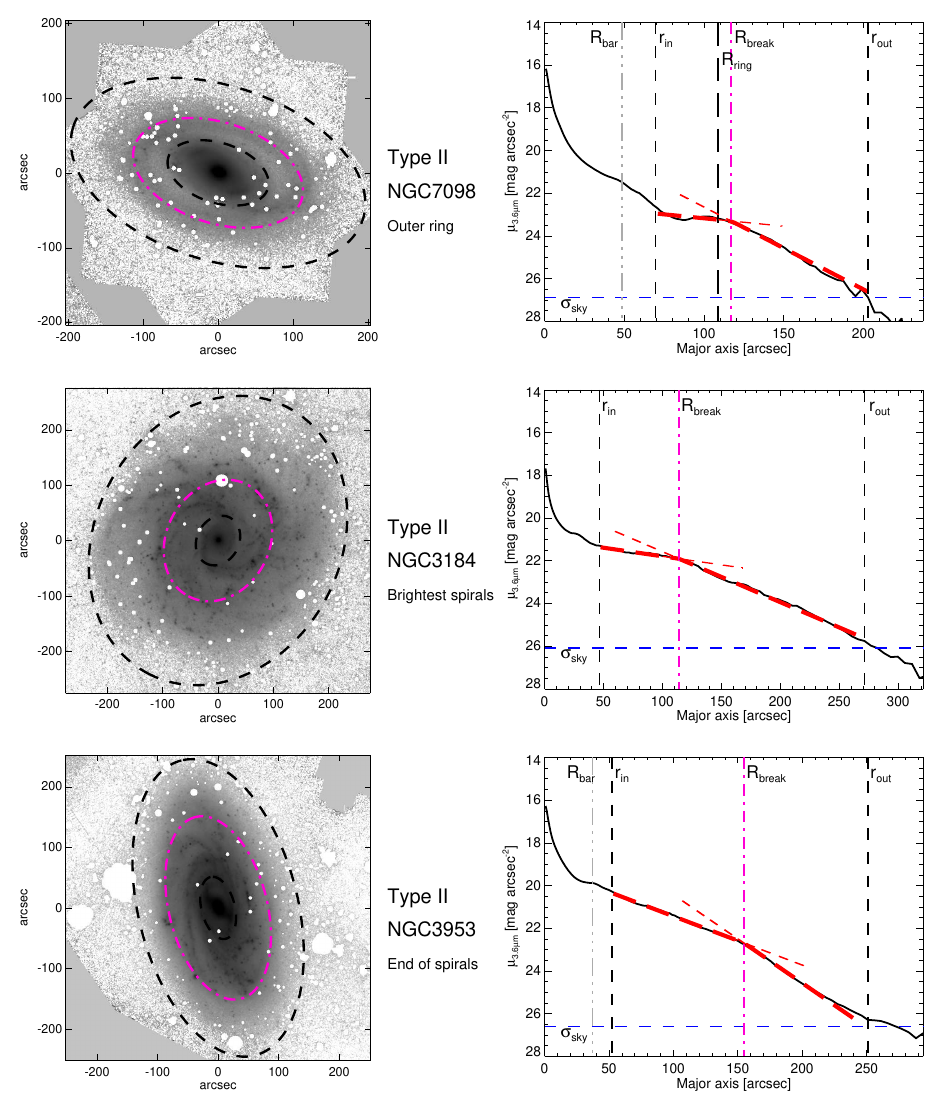}
    \caption{Same as figure \ref{prototypes}, but for type II examples only.}
    \label{prototypes_type2}
  \end{center}
\end{figure*}

During the fitting of the surface brightness profiles the galaxies were visually inspected and the connection of the break radii with the structural components were studied. In the identification of the structures we rely on the classifications given in  \citet{laurikainen2011} and Buta et al. (in preparation). The type II.i profiles are connected to bars, and we do not include further discussion of this type below.
				
\subsubsection{Type II breaks}
\label{type2connections}
                          
We found three distinct groups of type II profiles that cover $\sim 94 \%$ of that type: 1) breaks connected with inner and outer lenses, 2) breaks connected with outer rings or outer pseudorings, and 3) breaks connected with edge of the bright star formation regions in the spirals or apparent outer radius of the spirals. The last group is based on visual inspection of the images, with the break radii drawn over the galaxies, whereas the dimensions of the rings and lenses were taken from \citet{comeron2014} and \citet{laurikainen2011}. In barred galaxies only those rings and lenses with radii equal to or larger than the bar radii were considered, grouped as inner and outer rings/lenses, respectively. Nuclear rings/lenses were omitted.

Some type II breaks in the early-type disc galaxies appear to be associated with inner or outer lenses, and this was found in $\sim 8 \%$ (11 cases) of all type II profiles. In these galaxies the break radii coincide with the lens radii, as can be seen in Figure \ref{rings_breaks}, left panel. This connection is not surprising, since the lenses are defined as having a flat luminosity distributions with fairly sharp outer edges, which in turn appear as type II breaks in the radial surface brightness profiles.

We found that for the galaxies with type II profiles and an outer ring/pseudoring, these structures without exception are related to the breaks, with the ring radii coinciding with the break radii (Fig. \ref{rings_breaks} left panel, and Fig. \ref{prototypes_type2} upper panels for an example of such a galaxy). It appears that $\sim 48 \%$ (66 cases) of the type II profiles are associated either with an outer ring, an outer pseudoring, or an outer ringlens. If the profile types are then studied as a function of the  morphological type (Fig. \ref{rings_breaks_histo}) it is obvious that most of type II profiles in Hubble types $ -2 \lesssim T \lesssim 2$ are related to outer rings (see also Fig. \ref{break_distribution_simple} upper panel). In this Hubble type range type II profiles are particularly common in barred galaxies (Fig. \ref{break_distribution_simple}, lower middle panel).
                                          
In later type galaxies ($T \gtrsim 3$) the breaks are largely connected with the regions of the brightest star formation regions in the outer spiral arms (19 cases, $\sim 14\%$ of type II profiles, see Fig. \ref{prototypes_type2} middle panels), or with the apparent outermost radii of the spirals (34 cases, $\sim 24\%$ of type II profiles, Fig. \ref{prototypes_type2} lower panels). In the latter case, the observed outermost part of the surface brightness profile is usually a featureless disc (see Fig. \ref{prototypes_type2} lower panels). These two groups, associated with the spiral structures, form $\sim 38 \%$ of all type II breaks in our sample. These breaks are slightly less common in the barred than in the non-barred galaxies (see Fig. \ref{rings_breaks_histo} right panel).
              
Only for $\sim 6 \%$ (9 cases) of type II breaks no apparent morphological feature was connected with the break. \citet{erwin2005} proposed that some type II profiles could be caused by slight asymmetries, and indeed some of the cases in our study where no connection with morphological features was found could be due to this. After all, a majority of galaxies are asymmetric in the outskirts in some level (e.g. \citealt{zaritsky2013}).
                                           
In galaxies that have no outer structures, the inner rings are systematically slightly smaller than the break radii, but still correlate with each other. This is shown by the grey dashed line in Fig. \ref{rings_breaks} (left panel), which is a rough linear fit to the inner rings with a slope of 2.0. That both the inner and outer rings are correlated with the break radii is not unexpected if both rings are resonance structures. The outer rings are known to have a radius of around twice that of the inner rings (e.g. \citealt{kormendy1979}; \citealt{athanassoula1982}; \citealt{buta1986}; \citealt{sellwood1993}; \citealt{buta1996}; \citealt{rautiainen2000}; \citealt{comeron2014}).

\begin{figure*}		
  \begin{center}
    \includegraphics[width=0.9\linewidth]{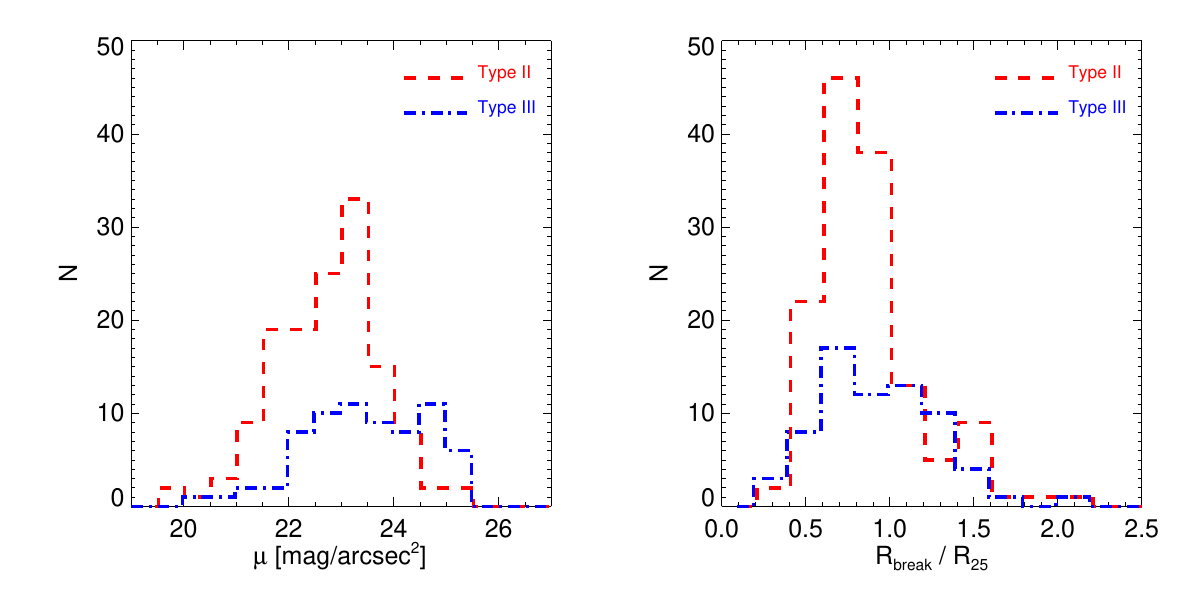}
    \caption{In the \textit{left panel} the distribution of surface brightness at the break radius of types II and III is shown, and in the \textit{right panel} the distributions of the break radius scaled with the B-band 25 mag arcsec$^{-2}$ isophotal level radius ($R_{break} / R_{25}$) is shown.}
    \label{break_surf_mag}

    \includegraphics[width=0.9\linewidth]{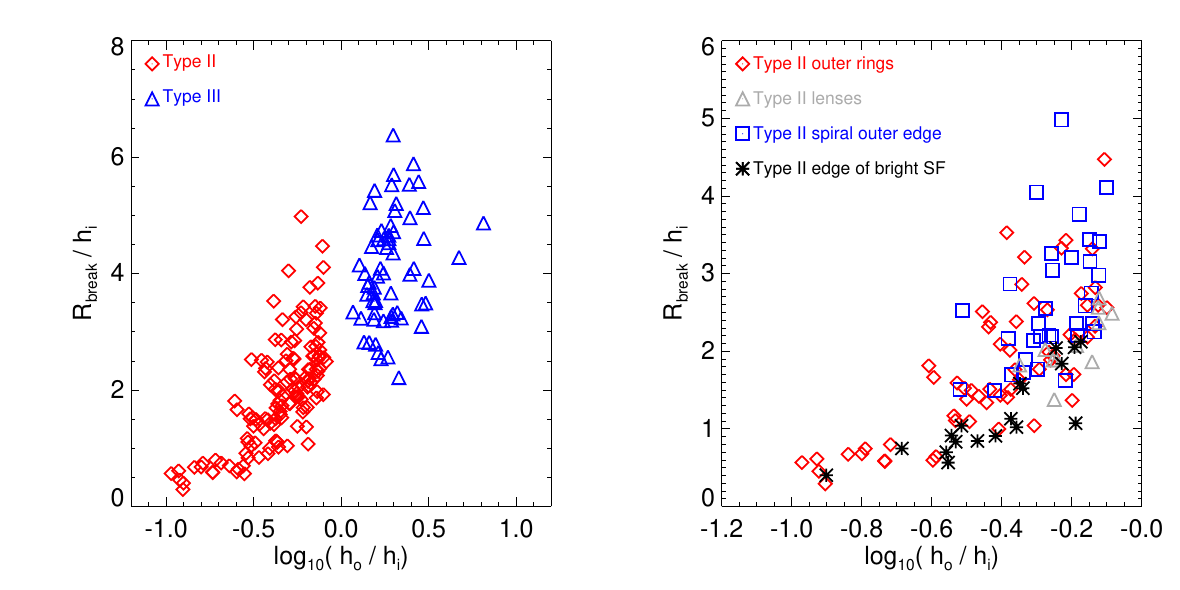}
    \caption{ The two parameters that have been used in literature to describe the breaks,  the ratio between the disc scalelengths inside and outside the break ($log_{10} (h_o / h_i)$), and the ratio between break radius and the scalelength of the disc inside the break ($R_{break} / h_i$). In the \emph{left panel} both type II and III profiles are shown, while in the \emph{right} panel only type II profiles are shown and different symbols are used to show with which structural component the break is associated with.}
    \label{break_strength}
  \end{center}
\end{figure*}

\subsubsection{Type III breaks}

We found that roughly 1/3 of all type III profiles in our sample can be directly connected with such structures as rings and lenses. However, in the majority of galaxies the type III features cannot be associated with any distinct structure. Also, independent of the component to which the type III profile is connected, the structures are 10 kpc or smaller in size, in distinction to the structures associated with type II profiles that extend up to $\sim 30$ kpc in size. Type III breaks found in the outer parts of galaxies most probably have some other explanation.

\begin{figure*}
  \begin{center}
        \includegraphics[width=0.75\textwidth]{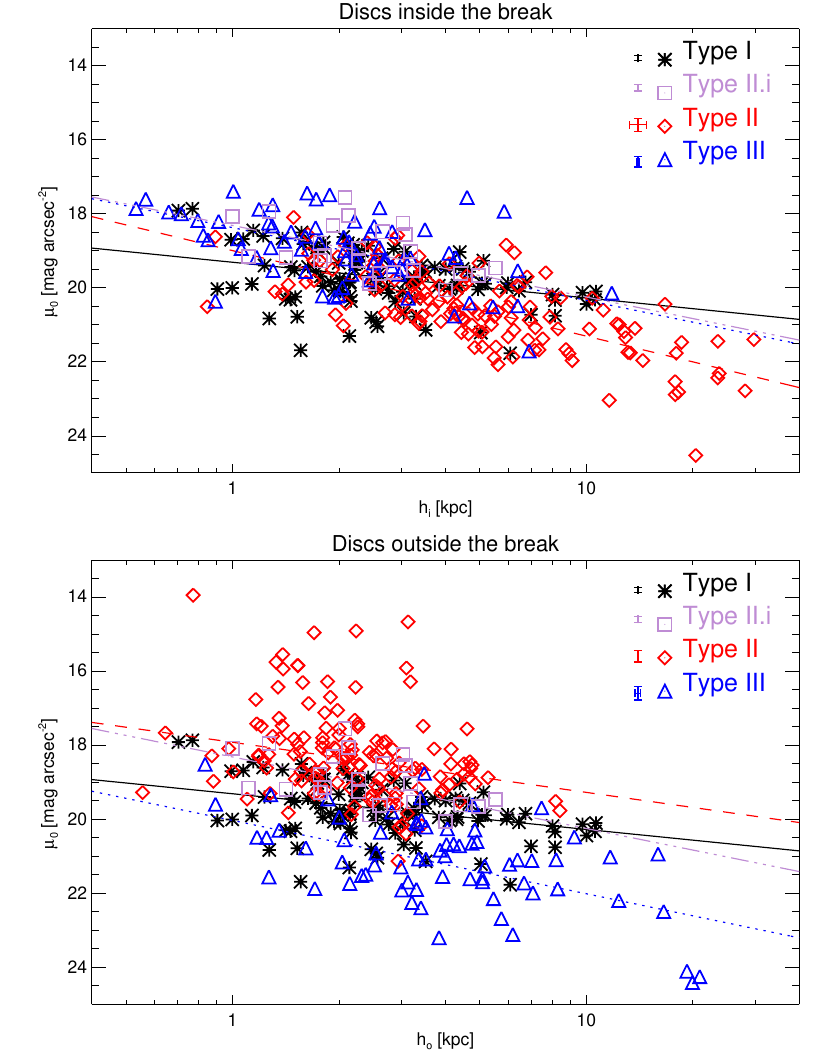}
    \caption{The inner disc scalelengths against the inner disc central surface brightness (\textit{top panel}), and the outer disc scalelengths against the outer disc central surface brightness (\textit{lower panel}). The points for type I and II.i profiles are the same in both panels. The lines show a fit to the points of the different profile types: \emph{solid line} for type I, \emph{triple-dotted dashed line} for type II.i, \emph{dashed line} for type II, and \emph{dotted line} for type III. The error bars show the mean uncertainties of the disc scalelengths and central surface brightnesses.}
    \label{hin_muin}
  \end{center}
\end{figure*} 	

\subsection{Parameters of the breaks}
\label{parameters-breaks}

The surface brightness at the break radius is one of the parameters determining the properties of the breaks. Generally type II and III breaks are found at similar surface brightnesses, although the median indicates that type III profiles appear at slightly lower surface brightnesses (see Fig. \ref{break_surf_mag} left panel). In addition, there is no large difference in the radius between type II and type III profiles, although the median $R_{break}/R_{25}$ is slightly larger for type III profiles (Fig. \ref{break_surf_mag} right panel).

Mainly two parameters have been used in the literature to characterise the break strengths in galaxies, the ratio of inner- and outer disc scalelengths around the break ($h_o/h_i$), and the ratio of break radius and the inner disc scalelength ($R_{break}/h_i$) (e.g. \citealt{vanderkruit1987}; \citealt{pohlen2006}; \citealt{maltby2012}; \citealt{martinnavarro2012}; \citealt{comeron2012}). The distributions of these parameters are shown in Figure \ref{break_strength} (left panel) for type II and III profiles, clearly distinguishing the two groups. It is worth noticing that type II profiles have a tail of low $R_{break}/h_i$ and $h_o/h_i$ ratio, indicating that the profile is flat inside the break.

We find that the flat inner discs are associated with outer rings or the outer regions of bright star formation regions in the spiral arms (Fig. \ref{break_strength} right panel). The breaks connected with the apparent outer edges of the spiral structures are also well separated from the breaks connected with strong star formation regions.

\begin{figure*}
  \begin{center}
    \includegraphics[width=0.9\textwidth]{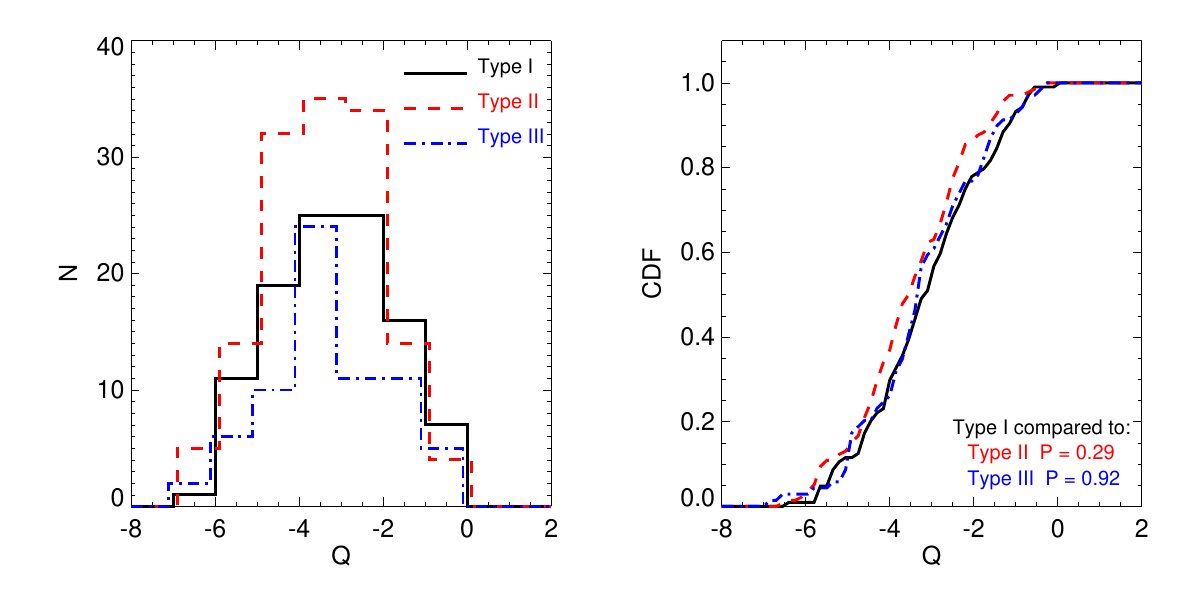}
    \caption{Distribution of the Dahari parameter. In the \textit{left panel} a histogram is shown, and in \textit{right panel} the cumulative distribution. The p-values are the probabilities from a 2-sided Kolmogorov--Smirnov test, indicating probabilities that the two compared samples are drawn from the same parent distribution.}
    \label{breaks_total_dahari}

    \includegraphics[width=0.9\textwidth]{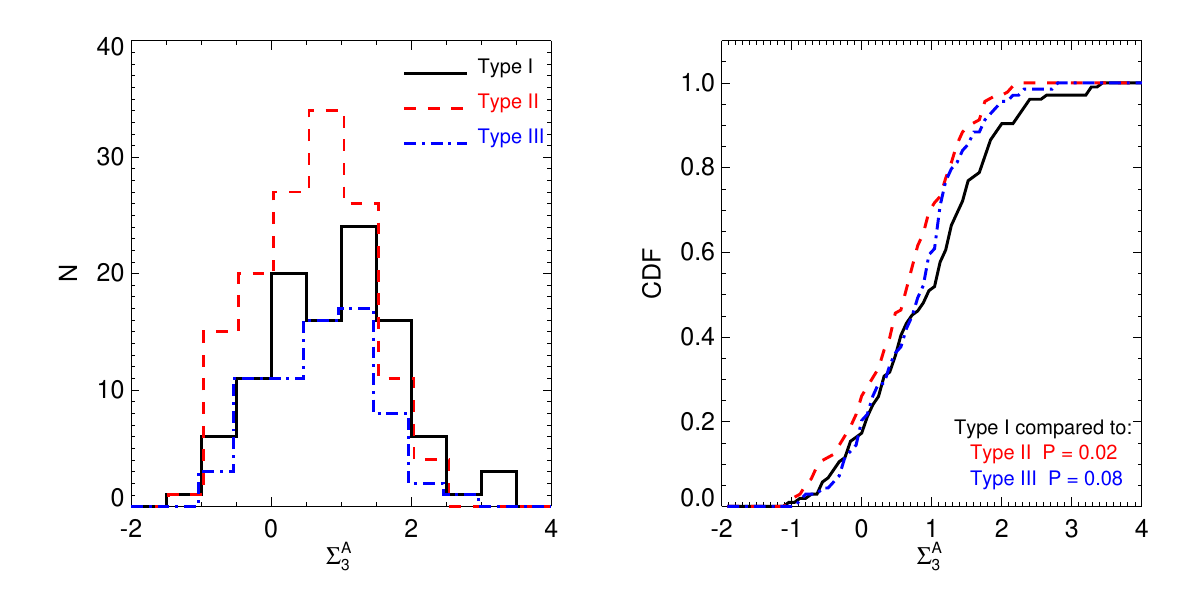}
    \caption{Same as Figure \ref{breaks_total_dahari}, but for the surface density of galaxies. Small $P=0.02$ for comparing between type I and II profiles indicate that they occur in statistically different environments as measured by $\Sigma_3^A$.}
    \label{breaks_total_surf}
  \end{center}
\end{figure*}
					
\subsection{Parameters of the discs}
\label{parameters-discs}

What makes the type II and III profiles deviate from the single exponential profile is still an open question. We show that the discs of type I galaxies follow the same scaling relation as the inner discs of type II and III galaxies, when the central surface brightness ($\mu_0$) is plotted against the scalelength of the disc ($h$) (Fig. \ref{hin_muin}, upper panel).  This scaling relation is similar to that obtained previously for the global disc parameters of S0s and bright spirals (\citealt{dejong1996}; \citealt{graham2001}; \citealt{graham2008}; \citealt{gadotti2009}; \citealt{laurikainen2010}). The inner discs of type II profiles extend to larger scalelength values than those of type III profiles. Concerning the disc parameters outside the break radius we see a large dispersion. This behaviour is largely related to the definition of the profile types, and particularly for type II profiles it might be a manifestation of several different origins of the break. Pure type I profiles are fairly similar to type II.i, but have on average higher extrapolated central surface brightness than type I profiles. We also compared the values of $\mu_0$ and $h$ of the inner- and outer discs individually for barred and non-barred galaxies for the main types, but found no differences. It is also worth noting that the estimated uncertainties are not large enough to affect the found scaling relations.


\section{Environmental properties}
\label{env_effects}

We found that the distributions of the environmental parameters, namely the surface density of the galaxies ($\Sigma_3^A$) and the Dahari parameter (Q) (Figs. \ref{breaks_total_dahari} and \ref{breaks_total_surf}), are fairly similar for all three disc profile types. The statistical tests (Kolmogorov--Smirnov $P$ and $D$) for both parameters are given in Table \ref{env-overall-stat}, where the mean values of the parameters for the profile types are also given. There is some hint in the parameter $\Sigma_3^A$, that type I profiles appear in denser galaxy environments than types II and III. This difference is statistically significant when type I and II profiles are compared (see Fig. \ref{breaks_total_surf} and Table \ref{env-overall-stat}).

\begin{figure*}
  \begin{center}
    \includegraphics[width=0.85\textwidth]{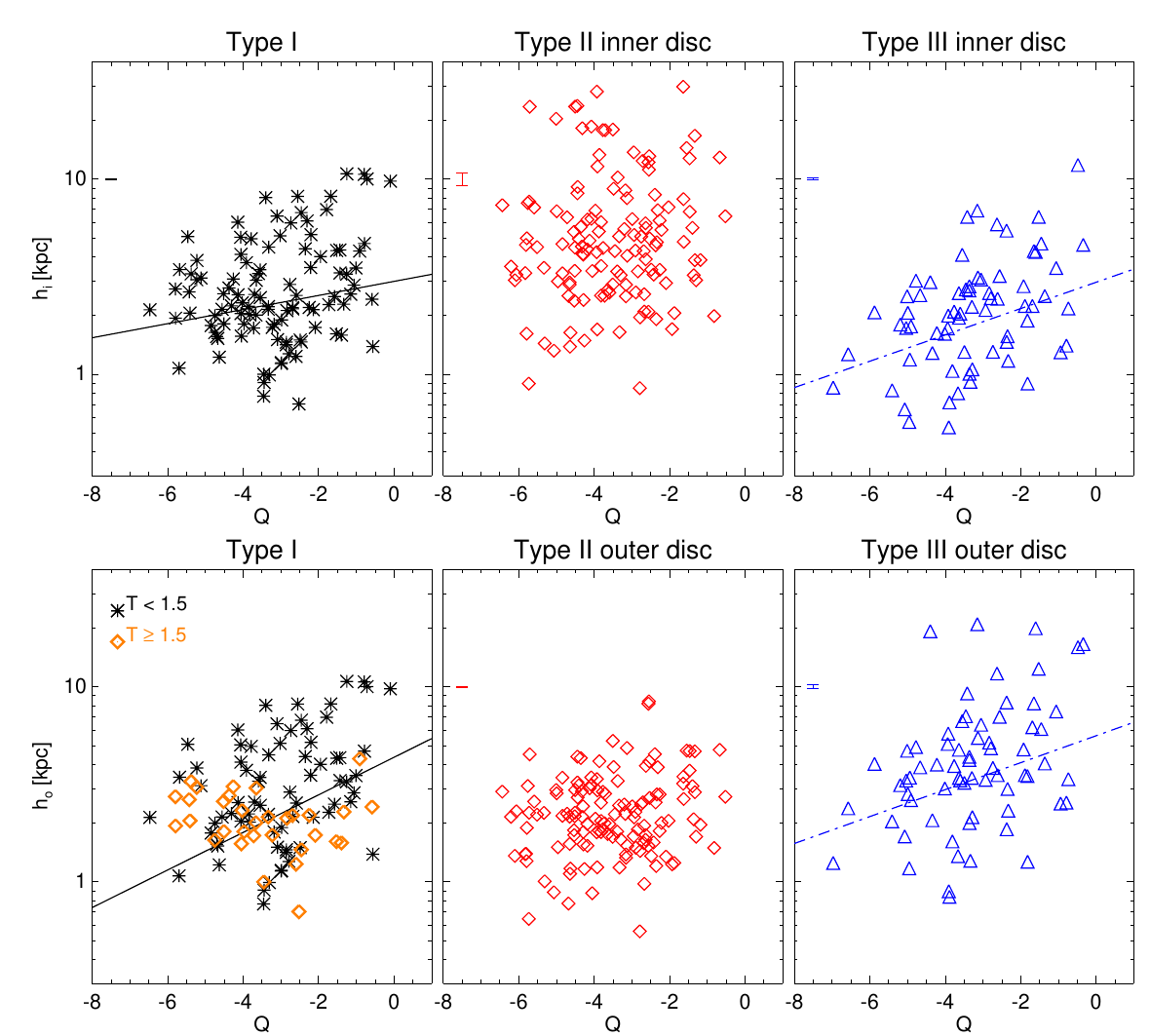}
    \caption{The disc scalelengths as a function of the Dahari parameter ($Q$) for type I profiles (\textit{left panels}), and the inner- (\textit{upper middle and right panels}) and outer (\textit{lower middle and right panels}) disc scalelengths as a function of the Dahari parameter ($Q$) for type II and III profiles. The correlations are statistically significant for type I discs ($P=0.026$), and for inner- and outer discs of type III ($P=0.0003$ and $P=0.0019$, respectively). The lines show a simple linear fit to the points. In the lower left panel the type I profiles are divided by Hubble type. The correlation for the early-type galaxies ($T< 1.5$) with type I disc is statistically significant ($P=0.003$), and the line shows a simple linear fit to the points in this bin. The error bars show the mean uncertainties of the disc scalelengths.}
    \label{dahari_outer}

    \includegraphics[width=0.85\textwidth]{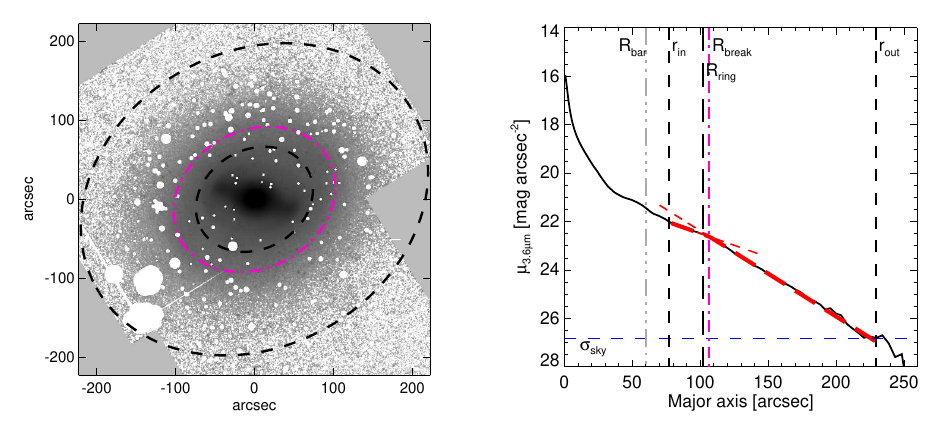}
    \caption{Image and radial surface brightness profile of NGC4596. Note the clear type II break that is associated with the outer ring.}
    \label{4596}
  \end{center}
\end{figure*}

Next we studied possible connections between the environmental properties and the various parameters describing the discs and breaks. Evidence that the environment affects the surface brightness profiles was found by looking at a possible correlation of the inner and outer disc scalelengths ($h_i$ and $h_o$) with the Dahari parameter. This correlation was found to be statistically significant for type III profiles inner and outer discs (Fig. \ref{dahari_outer} right upper and lower panels, Spearman's rank correlation test $P=0.0002$ and $P=0.0019$, respectively). The correlation is statistically significant also when $h$ is plotted against the Dahari parameter for type I profiles ($P=0.026$), although the scatter is very large. When this correlation is re-plotted separating the early-type galaxies ($T \le 1.5$ , Fig. \ref{dahari_outer} lower left panel), the scatter is significantly reduced ($P=0.003$). It is worth noting that no correlation is found in the disc or break parameters when using the surface density of the galaxies ($\Sigma_3^A$).
                
Our sample includes some visually identified interacting galaxies. Examples are NGC1097 (type II, $Q = -0.67$, $\Sigma_3^A = 0.72$), NGC0772 (type III, $Q = -1.06$, $\Sigma_3^A = 0.41$) and NGC5427 (Type II, $Q = -0.52$, $\Sigma_3^A = 0.20$). One galaxy, NGC3893 (Kar 302 A), from \citet{laurikainen2001} who studied M51-type galaxy pairs, is also in our sample. This galaxy has a type I profile and both environmental parameters are above the average ($Q = -1.39$, $\Sigma_3^A = 1.80$).

Previously \citet{pohlen2006} counted the number of neighbouring galaxies from SDSS within 1 Mpc projected radius. To select a likely companion they applied criteria for the recession velocity difference to the target galaxy and absolute magnitude ($| \Delta v | < 350$ km s$^{-1}$ and $M_{\text{r'}}< -16$ magnitude, respectively). They did not find a connection with the environment and the profile types. More recently \citet{maltby2012} has compared the outer disc scalelengths ($h_{o}$) and the break strengths ($log_{10} \, (h_{o} / h_{i}) $) between field and cluster spiral galaxies, at redshifts ($z_{phot} > 0.055$) higher than those in our sample ($z_{phot} \lesssim 0.020$). They focused only on breaks that appeared in the surface brightness profiles at $24.0 < \mu < 26.5$ mag arcsec$^{-2}$ in the V-band, which corresponds roughly $22.5 < \mu < 25.0$ mag arcsec$^{-2}$ at 3.6 $\mu m$. Compared to our Figure \ref{break_surf_mag} it means that they mostly missed type II and III breaks that appear at the surface brightnesses of $\mu_{3.6 \mu m} < 22.5$ mag arcsec$^{-2}$. They did not find any differences in the profile types between the field and cluster galaxies. 
		
\citet{erwin2012} and \cite{roediger2012} have argued that S0 galaxies in the Virgo cluster do not show any type II breaks at all. Our sample includes 24 galaxies that are in the Virgo cluster catalogue by \citet{binggeli1985}, of which 11 are S0s. Three of these galaxies have type II profile (see figure \ref{4596} for one example), and one galaxy has an outer type II, and an inner type III profile. The rest of the galaxies have four type I profiles, and three type III profiles. Clearly, from our small sample we can already say that type II profiles are not completely absent among the Virgo cluster S0 galaxies.

The environmental parameters were also examined separately for the main profile types in barred and non-barred galaxies, but no differences were found. The estimated uncertainties described in section \ref{proffunc} were not found to influence the found correlations. Type II.i profiles are rare and were not included in the environmental analysis.

\begin{table}
  \begin{center}
  \caption{The environmental parameters of the galaxies, as well as the Kolmogorov--Smirnov two tailed statistics test values. The parameter P describes the probability that the two compared distributions are drawn from the same initial distribution and the parameter D specifies the maximum deviation between the cumulative distributions. } 
    \label{env-overall-stat}
    \begin{tabular}{@{}l c c c}
      \multicolumn{4}{l}{Dahari parameter $Q$}	 \\
      \hline
      Type & Mean $\pm \, 1 \sigma$   &  &  \\
      \hline
      Type I & $-3.16 \pm 1.42$ &  &  \\
      Type II & $-3.53 \pm 1.37$  &  & \\
      Type III & $-3.26 \pm 1.44$ &  & \\
      \hline
      \multicolumn{4}{c}{ } \\
      \hline
      Type & Compared to & P & D \\
      \hline
      Type I & Type II & 0.29 & 0.15 \\
      Type I & Type III & 0.92 & 0.08 \\
      Type II & Type III & 0.39 & 0.13 \\
      \hline
      \multicolumn{4}{c}{ } \\
      
      \vspace*{20pt} \\
      
      \multicolumn{4}{l}{Surface density $\Sigma_3^A$}	 \\
      \hline
      Type & Mean $\pm \, 1 \sigma$ &  &  \\
      \hline
      Type I & $0.89 \pm 0.94 $ &  &  \\
      Type II & $0.56 \pm 0.77 $ & & \\
      Type III & $0.74 \pm 0.76$ & & \\
      \hline
      \multicolumn{4}{c}{ } \\
      \hline
      Type & Compared to & P & D \\
      \hline
      Type I & Type II & 0.02 & 0.20 \\
      Type I & Type III & 0.08 & 0.19 \\
      Type II & Type III & 0.22 & 0.15 \\
      \hline
    \end{tabular}
  \end{center}
\end{table}

\section{Discussion}
\label{discussion}

One of the big puzzles in the structure formation of galaxies has been, and still is, why a large majority of discs in galaxies are exponential. Were the galaxies formed like that or did that happen later by some internal or externally driven processes? In this study we examine the extent to which the observed deviations from the exponential profile can be associated with the observed morphological structures like bars, rings, and lenses. Such associations can be used to assess the presence of internal dynamical processes that can re-distribute mass. We also study in which way the galaxy environment is related to the profile type.

\subsection{Association to structure components of the galaxies}

\subsubsection{Type II}
\label{ii_dis_comp}

We found that almost all ($\sim 94 \%$) of type II breaks are associated with distinct morphological structures. In early-type disc galaxies ($T<3$) the breaks are typically connected with outer rings (outer ring/pseudo-ring/ringlens, $\sim 48\%$ of type II breaks) or lenses ($\sim 8\%$ of type II breaks). In later type galaxies the breaks are more likely associated with the outer edges of star formation regions, or with the apparent outer edges of the spirals (in total $\sim 38 \%$ of type II breaks). This leaves only $\sim 6 \%$ of type II breaks that could not be associated with any visible structure.
             
A possible connection between type II breaks and outer rings was proposed by \citet{pohlen2006}. We have systematically studied this connection, using existing ring size measurements from \citet{comeron2014} and \citet{laurikainen2011}. We find a strong correlation between the ring and break radii (Fig. \ref{rings_breaks}), suggesting that rings are causing the effect (see also appendix A in \citealt{kim2014}). This connection is expected if the outer rings are associated to the Outer Lindblad Resonance of the bar. The bar will effectively redistribute material and angular momentum in the disc thus forming the rings through the resonances, leading to deviations from a single exponential surface brightness profile. In this study outer lenses are also found to be connected with type II breaks in the early-type disc galaxies, in a similar manner as the outer rings (see Fig. \ref{rings_breaks}). This is natural if the outer lenses are also largely resonance structures, as suggested by \citet{laurikainen2013}. 
 
Additionally, we see a connection between the inner rings and type II breaks in those galaxies in which no outer rings or lenses are visible (see Fig. \ref{rings_breaks}). In those cases the breaks are nearly twice as large as the inner rings. This factor of two difference in size corresponds to that given in resonance theory for the inner and outer rings (e.g. \citealt{sellwood1993}; \citealt{buta1996}; \citealt{athanassoula2012}). It is possible that these galaxies have faint outer rings not apparent in the images, or that distinct outer rings have already disappeared.

In later type galaxies (see Fig. \ref{break_distribution_simple} lower middle panel), type II profiles are generally of a different nature and are connected with the star formation regions in spiral arms or with the apparent outer edges of the spiral structures (Fig. \ref{rings_breaks_histo}). While some of this change could be explained by a slightly lower bar fraction compared to early-type galaxies ($T < 3$), the bar fraction in type II profiles in late-type spirals is still as high as $40-50 \%$ (see Fig. \ref{break_distribution_simple} lower middle panel) and thus can not be the main factor causing it. Strong star formation in the spiral arms inside the break radius was found to be related to $\sim 14 \%$ of all the type II profiles. In these cases the spiral arms continue also outside the break radius. Star formation thresholds have been considered as one possible mechanism for type II break formation (see for example \citealt{schaye2004}). Breaks have been detected also in the radial surface brightness profiles based on H$\alpha$ emission line measurements of late-type disc galaxies, at $\sim 0.7 \, R_{25}$ \citep{christlein2010}, which coincide well with many of type II breaks of this study (Fig. \ref{break_surf_mag} right panel). Additionally, coupled bar-spiral resonances \citep{tagger1987} might be causing the observed breaks for some barred type II galaxies \citep{munozmateos2013}.
                 
Galaxies where the breaks are connected with the apparent outer radii of spiral structures comprise $\sim 24 \%$ of all type II profiles. In these cases the galaxy extends outwards from the break radius, but the outer region is featureless. This featureless region could consist of migrating old stars. Some evidence of migration is found in the colour profiles of galaxies (\citealt{azzollini2008}; \citealt{bakos2008}), where for type II profiles the discs outside the break get increasingly redder, indicating older stellar populations. In principle it is also possible that the outer discs are redder simply because of a change in the star formation profile of the disc, without any need for stellar radial migration \citep{sanchezblazquez2009}. Moreover, \citet{sanchezblazquez2009} claim that migration would remove the break from the stellar mass distribution. This is inconsistent with our results which show a similar fraction of type II breaks (see Table \ref{overall-stat}) in the near infrared (and thus in the stellar mass distribution), as is seen in the previous studies in the optical wavelengths (e.g. \citealt{gutierrez2011}). Simulations and star count observations of NGC7793 by \citet{radburnsmith2012} show that the ratio of the outer and inner disc scalelengths ($h_o/h_i$) increase with older stellar populations (i.e. the break gets smoother), as the oldest stars have had more time to migrate to radii where less in-situ star formation is expected. Thus, also the break in the stellar mass profile of the galaxy is smoothed, but they found that the break remains visible in the radial surface brightness profile of the old stellar population that matches the $3.6 \mu m$ profiles of our study. Similar behaviour with increasing stellar age was not seen in the simulations of \citet{sanchezblazquez2009} without taking migration into account. It is worth noting that for these type II profiles, associated with the outer edges of spiral structure, value of $R_{break}/h_i = 2.61 \pm 0.82$ is in agreement with the value of 2.6 derived from the simulation models for radial migration by \citet{roskar2008a}. However, the analysis of stellar populations is beyond the scope of this paper. 

\begin{figure*}
  \begin{center}
    \includegraphics[width=0.95\textwidth]{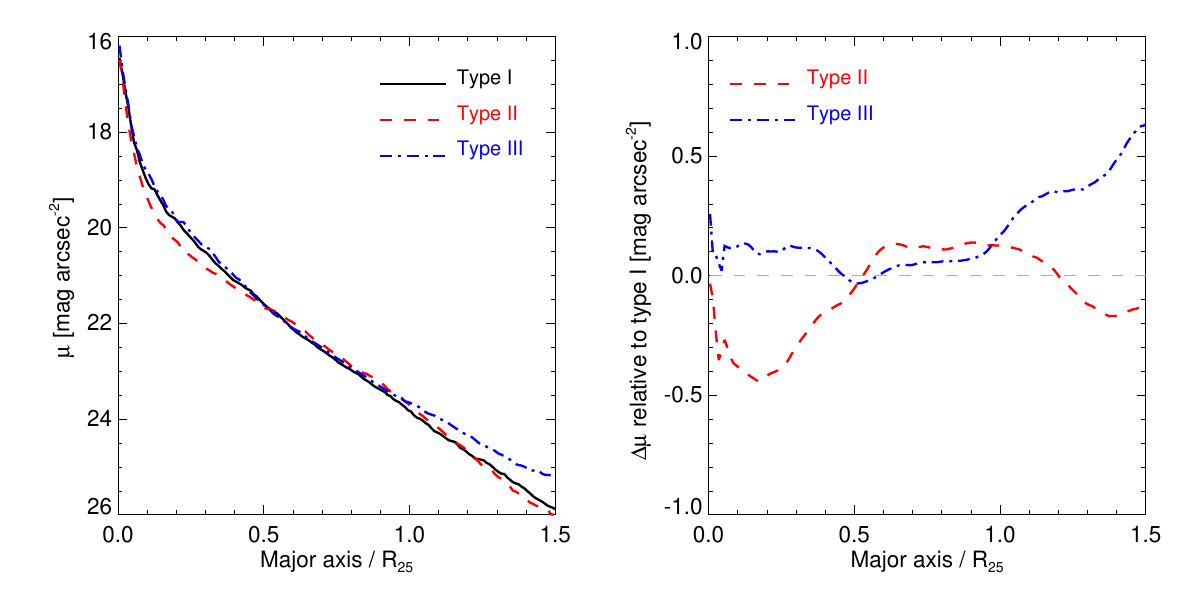}
    \caption{The median profiles of the 242 S$^4$G galaxies in the sample that have only one break in the surface brightness profile. In the \textit{left panel} the median profiles of type I, II, and III are shown, and in the \textit{right panel} type II and III median profiles are compared with type I median profile. The major axis has been scaled by the B-band 25 mag arcsec$^{-2}$ isophotal level radius.}
    \label{median_profiles}
  \end{center}
\end{figure*}

\subsubsection{Type III}

In approximately one third of type III profiles we could associate the break radius directly to galaxy structures like inner/outer lenses, or outer rings. Multiple exponential subsections associated with lenses have been previously studied by \citet{laurikainen2005} and \citet{laurikainen2009,laurikainen2011}, in the early-type disc galaxies. They showed that in such galaxies the exponential subsections appear at fairly high surface brightnesses. Our result fit into this picture as some of the type III breaks are indeed at relatively high surface brightnesses, and also because the type III breaks connected with the structural components appear at a fairly small radius (Figs. \ref{break_surf_mag} left panel, and \ref{rings_breaks} right panel).

Other structural components have also been proposed to create the type III profiles. \citet{bakos2012} found, using deep optical images for a small sample of face-on galaxies, that stellar halos systematically produce type III profiles at surface brightness levels of $\mu_{r'} \sim 28$ mag arcsec$^{-2}$. This surface brightness level corresponds roughly to a surface brightness of $\mu_{3.6 \mu m} \sim 27$ mag arcsec$^{-2}$, and is typically one magnitude fainter than the reach of our data. Nevertheless, bright stellar halos are seen in some galaxies in the S$^4$G sample. Additionally, the superposition of thin and thick discs has been proposed by \citet{comeron2012} to create some of the type III profiles. The type III profile would be formed when the thin disc has a lower scalelength than the thick disc.

In addition to the formation scenarios mentioned above, star formation in extended gas discs might also be involved, in at least some fraction of the type III breaks. The discovery of extended UV-discs, that continue far beyond the optical discs, indicates the presence of star formation at large galactocentric distances in about 25\% of disc galaxies (e.g. \citealt{gildepaz2005}; \citealt{thilker2005}; \citealt{zaritsky2007}). In some of the cases the extended UV-discs might be a result of galaxy interactions (for example M83, \citealt{gildepaz2007}), but often the galaxies are isolated.

\subsection{Scaling relations}
	
In Figure \ref{median_profiles} (right panel) we show the median surface brightness profiles of type II and III  normalized with the median type I profile, scaled with the B-band 25 mag arcsec$^{-2}$ isophotal level radius. It appears that type II profiles in the outer disc are closer to the single exponential discs. On the other hand, type III profiles in the inner disc are more similar to the single exponential discs. These differences in the profile types are seen also in the obtained scaling relations, although they mainly arise from the behaviour of the disc scalelengths. For example, in Fig. \ref{hin_muin} (upper panel) the type I and III profiles overlap in scalelength. In the same Fig. \ref{hin_muin} (lower panel) the types I and II overlap with each other in scalelength. The inner parts of type II profiles are flatter than the single exponential profiles. As discussed in the next section this phenomenon is most probably related to bar induced secular evolution in galaxies. Whether the extended outer discs of type III profiles are manifestations of environmental effect will be discussed in Section \ref{env-disc}.

\citet{gutierrez2011} compared the disc scalelength ($h$) and the central surface brightness ($\mu_0$) of the inner and outer parts of type II and III profiles, to the values of type I profiles. They divided their sample to early ($T\le3$) and late-type galaxies ($T>3$). Contrary to our study, they found that in both Hubble type bins the inner discs of type II profiles are similar to those of type I profiles. Also, in their study the outer parts of type II, and inner parts of type III profiles had shorter scalelengths and brighter $\mu_0$ than type I profiles. Using the same Hubble type bins we find that the inner disc scalelengths of type III are similar to type I profiles in both early and late-type galaxies (K--S test value $P=0.11$ in both bins). In late-type galaxies also the outer disc scalelengths of type II profiles are similar to type I discs ($P=0.28$). These differences between the studies could arise from the different wavelengths used in the studies.

The scatter among points in the $\mu_0$ vs. $h_o$ diagram (upper left corner in Fig. \ref{hin_muin}, lower panel) show no connection with the structural components of the galaxies or with the galaxy absolute magnitudes. Thus, the scatter could be related to various formation scenarios of the breaks. In principle it could be due to small fitting errors in the outermost parts of the discs that influence the extrapolated disc central surface brightnesses (see also \citealt{munozmateos2013}). However, our estimated uncertainties are not large enough to explain this scatter. 

\subsection{Bar related secular evolution of the discs}

Associating disc breaks with the different morphological structures, and considering the scaling relations discussed above, it is obvious that bars must play an important role in redistributing matter in galactic discs. Bars can transfer angular momentum among stars and therefore cause the disc to spread, while the gas in the disc is driven towards the bar resonances (for a recent review see \citealt{athanassoula2012}). In section \ref{ii_dis_comp} we discussed that nearly a half ($\sim 48$ \%) of type II profiles are connected with outer rings and that a correlation exists with the break radius and the inner ring radius, which are thought to be formed at the bar resonances. Further evidence of the influence of bars on the discs is seen in the bar fractions of the main profile types (see Table \ref{overall-stat}), and in the distributions of the breaks in barred and non-barred galaxies (see Fig. \ref{break_distribution_simple} lower panels). While types I and III are found equally in barred and non-barred galaxies ($\sim 50$ \% bar fraction), type II profiles are more common in barred galaxies ($\sim 72$ \% bar fraction). Most of the profiles in barred galaxies in Hubble types $T<3$ are of type II. And also, the majority of type II profile breaks are connected with the outer rings (see Fig. \ref{rings_breaks_histo}).

The outer profiles of type II are similar to those of types I and II.i, both among the barred and non-barred galaxies. In fact, the whole type II.i surface brightness profile is similar to that of type I, except inside the bar region where the surface brightness falls below the extrapolated disc. These intermediate cases between types I and II are caused by the bar on the underlying stellar disc, and similar profiles have been seen in N-body simulations (e.g. \citealt{athanassoula2002}; \citealt{valenzuela2003}).

Outer rings are not common in early-type S0 galaxies (S0$^o$, S0$^+$), where they seem to be replaced with lenses \citep{laurikainen2013}. Lenses in these galaxies appear in both barred and non-barred galaxies. It was discussed by \citet{laurikainen2013} that lenses in these galaxies might be largely structures formed in the earlier phase of galaxy evolution when the galaxies were still barred. We also find that the seven type II breaks (out of 66 type II breaks associated with outer rings) connected with outer rings, appear in non-barred galaxies. Formation of rings in non-barred galaxies is not well understood, but based on our study they are similarly connected with the type II breaks as in barred galaxies. 

It seems that bars in later type disc galaxies trigger the break formation in a different manner. We find that in Hubble types $T>3$ most of the type II breaks are connected with star formation in spiral arms, or with the apparent outer edges of the spirals (Fig. \ref{rings_breaks_histo}). These type II profiles appear mostly in the non-barred galaxies (Fig. \ref{break_distribution_simple} lower middle panel). 

We conclude that in the early-type disc galaxies ($T<3$) bar tends to flatten the disc profile inside the break radius via redistribution of matter, making them to deviate from type I profiles. On the other hand, in the late-type ($T>3$) galaxies the breaks are caused by star formation in the disc.

\subsection{Environmental effects}
\label{env-disc}

Minor-mergers in simulations are able to produce realistic type III profiles in galactic discs (e.g. \citealt{younger2007}; \citealt{elichemoral2011}). Major mergers are known to produce shells, loops, and tails, which in some cases can also be associated with type III profiles, in particular at low surface brightnesses (e.g. $\mu_v =26-29$ mag arcsec$^{-2}$, \citealt{janowiecki2010}). An example of a galaxy with shells in our sample is NGC0474 (see also \citealt{kim2012}), which shows a type III break in the disc associated with the shell structures.

Instead of looking for such direct evidence of galaxy interactions, we calculate the surface density of galaxies ($\Sigma_3^A$) and the Dahari parameter ($Q$), in order to evaluate possible environmental effects on galaxies.  For the early-type galaxies ($T<1.5$) with type I profile (Fig. \ref{dahari_outer} lower left panel) we find a statistically significant correlation between the disc scalelength ($h$) and the Dahari parameter ($Q$).  In principle this could be a manifestation of the well known morphology-density relation (e.g. \citealt{oemler1974}; \citealt{davis1976}; \citealt{dressler1980}): the early-type galaxies live in denser galaxy environments, where the tidal effects are more frequent. Early-type disc galaxies are also on average slightly brighter which could explain the larger scalelengths (e.g. \citealt{binggeli1988}; \citealt{courteau2007}). However, as no correlation was found between $h$ and the galaxy density parameter, $\Sigma_3^A$, it is more likely that the type I profiles are created in small galaxy groups where the tidal effects are efficient in modifying the discs. Alternatively they could be relics of major mergers. However, based on our study alone, it has not been ruled out that a majority of type I profiles were relics of initial disc formation, from the epoch where galaxy halos controlled the relation between the scalelength and galaxy brightness. 

For type III profiles we find that both the inner ($h_i$) and the outer ($h_o$) disc scalelength correlate with the Dahari parameter (Fig. \ref{dahari_outer} right panels): when the tidal effect increases, both scalelengths increase. The increase in $h_o$ with increasing tidal force is consistent with the picture in which the companion galaxies disturb the outer discs creating the observed type III profiles. What happens to the outer disc depends strongly on the orbital parameters of the encounter, the gas content, and the relative masses of the encountering galaxies (e.g. \citealt{laurikainen2001}; \citealt{younger2007}). More specifically, the type III profile could be a result of a tidal encounter when the main galaxy has a significant supply of gas, and the encounter occurs in a prograde orbit with moderate orbital angular momentum. In the above picture similar increase in $h_i$ of type III profiles with increasing tidal force, is less clear. The simulations of \citet{laurikainen2001} and \citet{younger2007} predict that in a minor merger $h_i$ remains similar to the initial disc before the encounter, increasing only $h_o$. On the other hand, it has been shown by \citet{elichemoral2011} that rotationally supported inner components (e.g. discs and rings) can be formed from the accreted material of the satellite galaxies. Therefore, it is possible that the observed properties of type III profiles could, at least partly, be triggered by tidal encounters in small galaxy groups. 

We did not find similar correlations between the surface density of the galaxies ($\Sigma_3^A$) and the parameters of the disc's breaks. It means that tidal interactions with the nearby companions are more likely to affect the discs than the surrounding galaxy density. This could also explain why in the previous studies of \citet{pohlen2006} and \citet{maltby2012} no connections were found with the profile types and the environment. They used more global measures of galaxy density (number of galaxies in an aperture, field/cluster comparison) that do not directly tell if the primary galaxy has a close companion. 
                 

\section{Summary and conclusions}
\label{sum-conclusion}

We present a detailed study of the disc surface brightness profiles of 248 galaxies using the 3.6 $\mu m$ images that form part of the Spitzer Survey of Stellar Structure in Galaxies (S$^4$G, \citealt{sheth2010}). Additionally, 80 galaxies were taken from the Near Infrared S0-Sa galaxy Survey (NIRS0S, \citealt{laurikainen2011}, observed at $K_{\text{s}}$-band). Using the radial surface brightness profiles we measured the properties of the main disc break type, first defined by \citet{erwin2005}. We associate the breaks with possible structural components in these galaxies using existing size measurements of rings and lenses. In addition, we carried out an environmental study of the sample galaxies using the 2 Micron All Sky Survey Extended Source Catalog (XSC), and the 2 Micron All Sky Survey Redshift Survey (RSC), and calculated the parameters describing the environmental galaxy density ($\Sigma_3^A$) and the Dahari parameter ($Q$) for the tidal interaction strength. 
	
Our main results are summarised as follows:

\begin{itemize}
  \item The fractions of the different profile types in the near infrared ($3.6 \, \mu m$ and $K_{\text{s}}$-band) are: type I $32 \pm 3$ \%, type II $42 \pm 3$ \%, and type III $21 \pm 2$ \%. We find also type II.i profiles in $7 \pm 2$ \% of the sample. In seven galaxies we see two breaks. These galaxies are counted twice and explain why the total percentage exceeds 100 \%.
  
  \item The inner parts of type III profiles are found to resemble the single exponential discs, while for type II it is the outer disc that more closely resemble the single exponential discs. This suggests that in galaxies with type II profile the evolution of the inner parts of the galaxy has been more significant, while in type III profiles the outer disc has gone through substantial evolution.
  
  \item $\sim 56 \%$ of type II profiles can be directly connected to outer lenses ($\sim 8 \%$), or to the outer rings, pseudorings, and ringlenses ($\sim 48 \%$). Almost all of the type II profiles with Hubble types $T<3$ are associated to these structures, with break radii that are coincident with the location of these structures. Therefore in galaxies of Hubble types $T<3$ the breaks are most likely associated to the resonances of bars.
  
  \item $\sim 38 \%$ of type II profiles can be visually connected to either the outer edges of intense star formation in the spiral arms or to the apparent outer radii of the spirals. These profiles appear mainly in Hubble types $T>3$, where they explain nearly all of type II profiles.
  
  \item Only approximately 1/3 of type III profiles could be associated with distinct morphological structures in the galaxies, such as lenses or outer rings. 
  
  \item For type III profiles a correlation was found between inner and outer disc scalelengths ($h_i$ and $h_o$) and the Dahari parameter ($Q$), indicating that nearby galaxy encounters are partly causing the upbending part of these profiles.

  \item The disc scalelengths ($h$) and central surface brightnesses ($\mu_0$) of the inner- and outer discs were found to be similar in barred and non-barred galaxies when the main types (I, II, III) are studied individually.

\end{itemize}

\section*{Acknowledgements}

We thank the referee for comments that have significantly improved the manuscript. The authors wish to thank the entire S$^4$G team for their efforts with this project. J.L. gratefully acknowledges financial support from the Vilho, Yrj\"o ja Kalle V\"ais\"al\"a foundation of the Finnish Academy of Science and Letters. J.L, E.L, H.S, and S.C acknowledge the support from Academy of Finland. E.A. and A.B. acknowledge the CNES (Centre National d'Etudes Spatiales - France) for financial support. We acknowledge financial support from the People Programme (Marie Curie Actions) of the European Union's FP7 2007-2013 to the DAGAL network under REA grant agreement number PITN-GA-2011-289313.

This research is based in part on observations made with the Spitzer Space Telescope, which is operated by the Jet Propulsion Laboratory, California Institute of Technology under a contract with NASA. We are grateful to the dedicated staff at the Spitzer Science Center for their help and support in planning and execution of this Exploration Science program. We also gratefully acknowledge support from NASA JPL/Spitzer grant RSA 1374189 provided for the S$^4$G project.

This research has made use of the NASA/IPAC Extragalactic Database (NED) which is operated by the Jet Propulsion Laboratory, California Institute of Technology, under contract with the National Aeronautics and Space Administration. This research has made use of SAOImage DS9, developed by Smithsonian Astrophysical Observatory.

\appendix
\section{Sample and disc profile parameters}
\label{app:sample}
\onecolumn

\begin{table}
\centering
\caption{General properties of the sample, and the environmental parameters. (1) Galaxy name, (2) the survey from which the galaxy is taken, (3) and (4) morphological types and numerical codes from \citet{laurikainen2011} for NIRS0S data and Buta et al. (in preparation) for S$^4$G data, (5) absolute blue magnitude from LEDA, (6) distance calculated from the 2MASS Redshift Survey recession velocity, (7) value of the Dahari parameter, (8) value of the surface density parameter, (9) projected radius at the distance of the galaxy used for the environmental analysis. The full table is available online.} 
\begin{tabular}{@{}c|c|c|c|c|c|c|c|c}
Galaxy & Survey & Morphological  &  T & M         & Distance & $Q$ & $\Sigma_3^A$ & $r$  \\
       &        & type    &    & [B-mag]   & [Mpc]    &     &              & [Mpc]\\
{\tiny (1)} & {\tiny (2)} & {\tiny (3)} & {\tiny (4)} & {\tiny (5)} & {\tiny (6)} & {\tiny (7)} & {\tiny (8)} & {\tiny (9)} \\
 \hline 
ESO137-010 & NIRS0S &  SA0$^+$:    & -1.0 & -21.84 & 46.92 & -0.78 & 1.50 & 1.\\
ESO337-010 & NIRS0S &  E$^+$3/SA0$^-$    & -3.0 & -21.45 & 81.22 & -1.67 & 0.92 & 1.\\
IC1438 & S$^4$G &  (R$_1$)SAB\underline{a}(r'l,nl)0/a  & 0.0 & -20.32 & 36.33 & -5.59 & -0.35 & 2.\\
IC1954 & S$^4$G &  SB(r\underline{s})cd  & 6.0 & -18.71 & 14.78 & -4.75 & 0.66 & 1.\\
IC2035 & S$^4$G &  SAB(l)0$^-$  & -3.0 & -18.98 & 20.75 & -4.95 & 0.20 & 1.\\
IC2051 & S$^4$G &  SB(\underline{r}s)b  & 3.0 & -20.25 & 24.35 & -5.80 & -0.96 & 3.\\
IC2627 & S$^4$G &  SA(s)bc  & 4.0 & -20.41 & 28.96 & -4.51 & 0.15 & 1.\\
IC2764 & S$^4$G &  (R)SA(l)0$^+$  & -1.0 & -18.98 & 23.32 & -4.84 & 0.04 & 1.\\
IC3102 & S$^4$G &  (R'L)SAB(\underline{r}s)0/a  & 0.0 & -19.73 & 30.89 & -2.75 & 1.16 & 1.\\
IC4214 & S$^4$G &  (R$_1$)S\underline{A}B$_{\text{a}}$(r'l,nr,nb)0/a  & 0.0 & -20.78 & 32.08 & -4.06 & -0.36 & 2.\\
IC4329 & NIRS0S &  SA0$^o$  shells/ripples    & -2.0 & -22.11 & 62.74 & -0.48 & 1.77 & 1.\\
IC4991 & NIRS0S &  coreE    & -3.0 & -22.16 & 78.17 & -0.78 & 1.26 & 1.\\
IC5240 & S$^4$G &  SB$_{\text{x}}$(r)0/a  & 0.0 & -19.35 & 24.26 & -4.38 & -0.01 & 1.\\
IC5267 & S$^4$G &  (RL)SA(r,l,nl)0/a  & 0.0 & -20.44 & 23.46 & -2.56 & 0.95 & 1.\\
IC5273 & S$^4$G &  SB(s)c  & 5.0 & -18.94 & 17.85 & -3.49 & 0.89 & 1.\\
IC5325 & S$^4$G &  SA(s)bc  & 4.0 & -19.75 & 20.89 & -4.50 & 0.26 & 1.\\
IC5328 & NIRS0S &  S$\underline{\rm  A}$B$_a$(l)0$^-$    & -3.0 & -21.00 & 42.79 & -5.46 & -0.60 & 2.\\
NGC0063 & S$^4$G &  S\underline{A}B(s)0/a  & 0.0 & -19.04 & 16.11 & -5.73 & 0.08 & 1.\\
NGC0150 & S$^4$G &  SAB(r\underline{s})ab  & 2.0 & -20.26 & 21.31 & -5.79 & -0.46 & 2.\\
NGC0157 & S$^4$G &  SA(s)bc  & 4.0 & -21.41 & 23.40 & -5.72 & -0.66 & 3.\\
NGC0210 & S$^4$G &  (R$_1$',R2'L)SAB(r'l,nr)ab  & 2.0 & -20.46 & 22.29 & -3.71 & 0.46 & 1.\\
NGC0254 & S$^4$G &  (R)SAB(l)0$^+$  & -1.0 & -19.17 & 21.18 & -4.51 & -0.14 & 2.\\
NGC0470 & S$^4$G &  S\underline{A}B(r\underline{s})ab  & 2.0 & -20.65 & 32.97 & -1.60 & -0.10 & 2.\\
NGC0474 & S$^4$G &  (R')SAB(l)0/a (shells/ripples)  & 0.0 & -20.46 & 32.29 & -1.60 & -0.07 & 2.\\
NGC0484 & NIRS0S &  SA0$^-$    & -3.0 & -21.18 & 70.90 & -3.32 & 0.19 & 1.\\
NGC0507 & NIRS0S &  (L)SAB0$^-$    & -3.0 & -22.10 & 68.53 & -0.09 & 2.17 & 1.\\
NGC0524 & NIRS0S &  (L)SA(l,nl)0$^o$    & -2.0 & -21.72 & 33.04 & -1.82 & 1.69 & 1.\\
NGC0578 & S$^4$G &  SB(s)cd  & 6.0 & -20.59 & 22.61 & -2.26 & -0.73 & 3.\\
NGC0584 & S$^4$G &  S\underline{A}B0$^-$  & -3.0 & -20.95 & 26.00 & -1.01 & 1.46 & 1.\\
NGC0613 & S$^4$G &  SB(\underline{r}s,nr)b  & 3.0 & -20.88 & 20.57 & -5.43 & -0.51 & 2.\\
NGC0628 & S$^4$G &  SA(s)c  & 5.0 & -20.72 & 9.12 & -4.84 & 0.71 & 1.\\
NGC0680 & S$^4$G &  SA(l)0$^o$ pec  & -2.0 & -20.58 & 38.89 & -1.89 & 1.43 & 1.\\
NGC0701 & S$^4$G &  SB(\underline{r}s)d  & 7.0 & -19.84 & 25.43 & -3.97 & 0.81 & 1.\\
NGC0718 & S$^4$G &  (R')SAB(rs,nl)a  & 1.0 & -19.65 & 24.07 & -5.51 & -0.98 & 4.\\
NGC0772 & S$^4$G &  SA(l,s)ab pec  & 2.0 & -22.81 & 34.33 & -1.06 & 0.41 & 1.\\
NGC0864 & S$^4$G &  SA\underline{B}(r\underline{s})bc  & 4.0 & -20.62 & 21.69 & -6.06 & -0.71 & 3.\\
NGC0890 & NIRS0S &  SA0$^-$    & -3.0 & -21.39 & 55.25 & -3.02 & -0.17 & 2.\\
NGC0918 & S$^4$G &  SAB(s)c  & 5.0 & -20.65 & 20.93 & -5.33 & -0.60 & 2.\\
NGC0936 & S$^4$G &  (L)SB(\underline{r}s,bl)0$^+$  & -1.0 & -20.37 & 18.61 & -2.37 & 1.23 & 1.\\
NGC0986 & S$^4$G &  (R')SB(rs,nr:,nb)ab  & 2.0 & -20.62 & 26.85 & -6.47 & -1.06 & 4.\\
$\dots$ & & & & & & & & \\
\hline
\end{tabular}  
\label{app:a}
\end{table}

\begin{landscape}
\begin{table}
\centering
\caption{Disc-profile parameters for the studied galaxies. (1) Galaxy name, (2) disc-profile type, (3) and (4) break radius in arcseconds and kiloparsecs, (5) surface brightness at the break in 3.6 $\mu m$ AB magnitude, (6) the extrapolated central surface brightness of the disc inside the break in 3.6 $\mu m$ AB magnitude, (7) and (8) scalelength of the disc inside the break in arcseconds and kiloparsecs (for type I this is the main disc scalelength), (9) the extrapolated central surface brightness of the disc outside the break in 3.6 $\mu m$ AB magnitude, (10) and (9) scalelength of the disc outside the break in arcseconds and kiloparsecs.The full table is available online. }
\begin{tabular}{@{} c | c | c | c | c | c | c | c | c | c | c }
Galaxy & Prof. & \multicolumn{2}{c|}{R$_{break}$} & $\mu_{break}$                         & $\mu_{i}$& \multicolumn{2}{c|}{h$_{i}$} & $\mu_{o}$ & \multicolumn{2}{c}{h$_{o}$} \\
       & type  & [kpc]             & [''] & $\left[ \frac{mag}{arcsec^2} \right]$ & $\left[ \frac{mag}{arcsec^2} \right]$ & [kpc]       & [''] & $\left[ \frac{mag}{arcsec^2} \right]$ & [kpc]       & [''] \\
{\tiny (1)} & {\tiny (2)} & {\tiny (3)} & {\tiny (4)} & {\tiny (5)} & {\tiny (6)} & {\tiny (7)} & {\tiny (8)} & {\tiny (9)} & {\tiny (10)} & {\tiny (11)} \\
 \hline
ESO137-010 & I & ------ & ------ & ------ & 19.57 $\pm$ 0.00 & 4.67 $\pm$ 0.00 & 20.55 $\pm$ 0.00 & ------ & ------ & ------ \\
ESO337-010 & I & ------ & ------ & ------ & 20.76 $\pm$ 0.00 & 8.16 $\pm$ 0.00 & 20.73 $\pm$ 0.00 & ------ & ------ & ------ \\
IC1438 & II & 9.55 $\pm$ 0.17 & 54.21 $\pm$ 0.95 & 23.22 $\pm$ 0.07 & 21.69 $\pm$ 0.10 & 7.15 $\pm$ 0.58 & 40.58 $\pm$ 3.28 & 19.14 $\pm$ 0.06 & 2.58 $\pm$ 0.03 & 14.63 $\pm$ 0.18 \\
IC1954 & I & ------ & ------ & ------ & 19.44 $\pm$ 0.04 & 1.64 $\pm$ 0.01 & 22.82 $\pm$ 0.20 & ------ & ------ & ------ \\
IC2035 & III & 2.96 $\pm$ 0.07 & 29.40 $\pm$ 0.66 & 23.08 $\pm$ 0.11 & 17.61 $\pm$ 0.08 & 0.57 $\pm$ 0.01 & 5.65 $\pm$ 0.12 & 20.49 $\pm$ 0.11 & 1.17 $\pm$ 0.04 & 11.66 $\pm$ 0.37 \\
IC2051 & II & 6.92 $\pm$ 0.12 & 58.65 $\pm$ 1.03 & 21.36 $\pm$ 0.06 & 19.68 $\pm$ 0.34 & 4.60 $\pm$ 1.88 & 38.95 $\pm$ 15.92 & 15.94 $\pm$ 0.07 & 1.39 $\pm$ 0.02 & 11.78 $\pm$ 0.13 \\
IC2627 & I & ------ & ------ & ------ & 19.63 $\pm$ 0.05 & 2.58 $\pm$ 0.06 & 18.40 $\pm$ 0.39 & ------ & ------ & ------ \\
IC2764 & I & ------ & ------ & ------ & 19.69 $\pm$ 0.06 & 1.78 $\pm$ 0.05 & 15.71 $\pm$ 0.44 & ------ & ------ & ------ \\
IC3102 & II & 13.28 $\pm$ 0.12 & 88.65 $\pm$ 0.82 & 24.06 $\pm$ 0.04 & 21.20 $\pm$ 0.14 & 5.08 $\pm$ 0.32 & 33.90 $\pm$ 2.16 & 18.28 $\pm$ 0.09 & 2.50 $\pm$ 0.04 & 16.69 $\pm$ 0.25 \\
IC4214 & I & ------ & ------ & ------ & 20.03 $\pm$ 0.06 & 4.11 $\pm$ 0.09 & 26.41 $\pm$ 0.57 & ------ & ------ & ------ \\
IC4329 & III & 33.36 $\pm$ 0.91 & 109.68 $\pm$ 2.99 & 23.22 $\pm$ 0.09 & 20.14 $\pm$ 0.03 & 11.81 $\pm$ 0.18 & 38.83 $\pm$ 0.60 & 20.94 $\pm$ 0.02 & 15.95 $\pm$ 0.11 & 52.45 $\pm$ 0.36 \\
IC4991 & I & ------ & ------ & ------ & 20.33 $\pm$ 0.08 & 10.58 $\pm$ 0.30 & 27.91 $\pm$ 0.79 & ------ & ------ & ------ \\
IC5240 & II.i & ------ & ------ & ------ & 19.26 $\pm$ 0.15 & 2.28 $\pm$ 0.07 & 19.35 $\pm$ 0.62 & ------ & ------ & ------ \\
IC5267 & III & 10.30 $\pm$ 0.69 & 90.59 $\pm$ 6.03 & 22.57 $\pm$ 0.19 & 19.20 $\pm$ 0.24 & 3.18 $\pm$ 0.32 & 27.99 $\pm$ 2.84 & 21.11 $\pm$ 0.08 & 7.02 $\pm$ 0.20 & 61.68 $\pm$ 1.77 \\
IC5273 & II & 6.78 $\pm$ 0.06 & 78.30 $\pm$ 0.74 & 23.48 $\pm$ 0.06 & 19.27 $\pm$ 0.02 & 1.77 $\pm$ 0.01 & 20.41 $\pm$ 0.17 & 17.76 $\pm$ 0.01 & 1.30 $\pm$ 0.00 & 15.01 $\pm$ 0.03 \\
       & III & 9.18 $\pm$ 0.02 & 106.11 $\pm$ 0.21 & 25.32 $\pm$ 0.00 & 17.76 $\pm$ 0.01 & 1.30 $\pm$ 0.00 & 15.01 $\pm$ 0.03 & 22.25 $\pm$ 0.07 & 3.21 $\pm$ 0.07 & 37.15 $\pm$ 0.77 \\
IC5325 & I & ------ & ------ & ------ & 18.93 $\pm$ 0.01 & 1.81 $\pm$ 0.00 & 17.89 $\pm$ 0.04 & ------ & ------ & ------ \\
IC5328 & I & ------ & ------ & ------ & 19.88 $\pm$ 0.04 & 5.07 $\pm$ 0.08 & 24.44 $\pm$ 0.38 & ------ & ------ & ------ \\
NGC0063 & II & 2.11 $\pm$ 0.12 & 26.96 $\pm$ 1.49 & 21.23 $\pm$ 0.17 & 18.62 $\pm$ 0.07 & 0.89 $\pm$ 0.04 & 11.45 $\pm$ 0.53 & 17.67 $\pm$ 0.06 & 0.65 $\pm$ 0.01 & 8.26 $\pm$ 0.10 \\
NGC0150 & I & ------ & ------ & ------ & 19.96 $\pm$ 0.08 & 2.73 $\pm$ 0.05 & 26.45 $\pm$ 0.52 & ------ & ------ & ------ \\
NGC0157 & II & 6.40 $\pm$ 0.06 & 56.40 $\pm$ 0.49 & 20.50 $\pm$ 0.03 & 19.59 $\pm$ 0.05 & 7.68 $\pm$ 0.60 & 67.68 $\pm$ 5.29 & 17.42 $\pm$ 0.01 & 2.26 $\pm$ 0.00 & 19.91 $\pm$ 0.04 \\
NGC0210 & II & 12.11 $\pm$ 0.06 & 112.02 $\pm$ 0.57 & 23.43 $\pm$ 0.04 & 22.54 $\pm$ 0.02 & 17.83 $\pm$ 0.47 & 165.00 $\pm$ 4.30 & 18.64 $\pm$ 0.11 & 2.83 $\pm$ 0.06 & 26.17 $\pm$ 0.52 \\
NGC0254 & II & 5.76 $\pm$ 0.14 & 56.10 $\pm$ 1.36 & 22.75 $\pm$ 0.06 & 20.91 $\pm$ 0.18 & 3.39 $\pm$ 0.46 & 32.99 $\pm$ 4.50 & 19.87 $\pm$ 0.01 & 2.17 $\pm$ 0.00 & 21.11 $\pm$ 0.04 \\
NGC0470 & II.i & ------ & ------ & ------ & 17.56 $\pm$ 0.14 & 2.08 $\pm$ 0.06 & 13.01 $\pm$ 0.37 & ------ & ------ & ------ \\
NGC0474 & III & 18.11 $\pm$ 0.15 & 115.71 $\pm$ 0.98 & 25.40 $\pm$ 0.03 & 20.76 $\pm$ 0.01 & 4.24 $\pm$ 0.02 & 27.05 $\pm$ 0.10 & 24.41 $\pm$ 0.33 & 19.96 $\pm$ 3.56 & 127.50 $\pm$ 22.73 \\
NGC0484 & I & ------ & ------ & ------ & 19.49 $\pm$ 0.07 & 4.47 $\pm$ 0.10 & 13.02 $\pm$ 0.29 & ------ & ------ & ------ \\
NGC0507 & I & ------ & ------ & ------ & 20.13 $\pm$ 0.11 & 9.76 $\pm$ 0.55 & 29.38 $\pm$ 1.65 & ------ & ------ & ------ \\
NGC0524 & III & 4.83 $\pm$ 1.56 & 30.15 $\pm$ 9.72 & 20.22 $\pm$ 0.91 & 17.49 $\pm$ 0.65 & 1.88 $\pm$ 0.56 & 11.75 $\pm$ 3.52 & 18.76 $\pm$ 0.45 & 3.49 $\pm$ 0.52 & 21.76 $\pm$ 3.23 \\
NGC0578 & II & 9.34 $\pm$ 0.07 & 85.20 $\pm$ 0.65 & 22.52 $\pm$ 0.03 & 20.82 $\pm$ 0.10 & 6.13 $\pm$ 0.52 & 55.94 $\pm$ 4.76 & 18.86 $\pm$ 0.02 & 2.81 $\pm$ 0.01 & 25.59 $\pm$ 0.10 \\
NGC0584 & I & ------ & ------ & ------ & 19.56 $\pm$ 0.07 & 3.50 $\pm$ 0.08 & 27.78 $\pm$ 0.62 & ------ & ------ & ------ \\
NGC0613 & II.i & ------ & ------ & ------ & 18.26 $\pm$ 0.21 & 3.03 $\pm$ 0.11 & 30.42 $\pm$ 1.09 & ------ & ------ & ------ \\
NGC0628 & II & 12.39 $\pm$ 0.53 & 280.01 $\pm$ 12.09 & 24.24 $\pm$ 0.21 & 20.20 $\pm$ 0.04 & 3.29 $\pm$ 0.06 & 74.38 $\pm$ 1.36 & 18.12 $\pm$ 0.38 & 2.18 $\pm$ 0.10 & 49.26 $\pm$ 2.35 \\
NGC0680 & III & 6.24 $\pm$ 0.48 & 33.08 $\pm$ 2.55 & 22.09 $\pm$ 0.22 & 19.11 $\pm$ 0.13 & 2.24 $\pm$ 0.14 & 11.87 $\pm$ 0.72 & 20.21 $\pm$ 0.09 & 3.52 $\pm$ 0.10 & 18.67 $\pm$ 0.51 \\
NGC0701 & I & ------ & ------ & ------ & 18.78 $\pm$ 0.04 & 1.81 $\pm$ 0.02 & 14.65 $\pm$ 0.20 & ------ & ------ & ------ \\
NGC0718 & II & 4.67 $\pm$ 0.16 & 40.01 $\pm$ 1.41 & 21.95 $\pm$ 0.04 & 20.82 $\pm$ 0.32 & 4.48 $\pm$ 4.72 & 38.41 $\pm$ 40.45 & 19.66 $\pm$ 0.01 & 2.21 $\pm$ 0.01 & 18.93 $\pm$ 0.06 \\
NGC0772 & III & 7.74 $\pm$ 1.61 & 46.51 $\pm$ 9.68 & 20.74 $\pm$ 0.50 & 18.43 $\pm$ 0.33 & 3.50 $\pm$ 0.68 & 21.02 $\pm$ 4.06 & 19.69 $\pm$ 0.28 & 7.47 $\pm$ 0.82 & 44.90 $\pm$ 4.92 \\
NGC0864 & II & 10.18 $\pm$ 0.69 & 96.77 $\pm$ 6.58 & 23.38 $\pm$ 0.24 & 19.94 $\pm$ 0.04 & 3.23 $\pm$ 0.07 & 30.72 $\pm$ 0.63 & 18.55 $\pm$ 0.24 & 2.30 $\pm$ 0.11 & 21.85 $\pm$ 1.09 \\
NGC0890 & I & ------ & ------ & ------ & 19.27 $\pm$ 0.04 & 5.12 $\pm$ 0.08 & 19.11 $\pm$ 0.30 & ------ & ------ & ------ \\
NGC0918 & II & 7.20 $\pm$ 0.12 & 70.91 $\pm$ 1.21 & 22.39 $\pm$ 0.04 & 20.15 $\pm$ 0.03 & 3.51 $\pm$ 0.08 & 34.60 $\pm$ 0.78 & 18.92 $\pm$ 0.02 & 2.26 $\pm$ 0.01 & 22.24 $\pm$ 0.13 \\
NGC0936 & II & 8.63 $\pm$ 0.08 & 95.63 $\pm$ 0.85 & 21.99 $\pm$ 0.03 & 19.89 $\pm$ 0.05 & 4.57 $\pm$ 0.13 & 50.70 $\pm$ 1.48 & 18.26 $\pm$ 0.01 & 2.55 $\pm$ 0.01 & 28.27 $\pm$ 0.08 \\
NGC0986 & II.i & ------ & ------ & ------ & 19.03 $\pm$ 0.06 & 3.19 $\pm$ 0.04 & 24.53 $\pm$ 0.33 & ------ & ------ & ------ \\
$\dots$ & & & & & & & & & & \\
\hline
\end{tabular}  
\label{app:b}
\end{table}
\end{landscape}

\bsp

\label{lastpage}

\end{document}